\documentclass[sigconf]{acmart}

\usepackage{gensymb}
\usepackage{microtype}                 % use micro-typography (slightly more compact, better to read)
\usepackage{tabu}                      % only used for the table example
\usepackage{booktabs}                  % only used for the table example
\usepackage{xcolor}
\usepackage{color}
\usepackage{enumerate}
\usepackage{enumitem}
\usepackage{textcomp}
\newcommand{\revision}[1] 
{\textcolor{black}{#1}}

\AtBeginDocument{%
  }

\setcopyright{acmcopyright}
\copyrightyear{2018}
\acmYear{2018}
\acmDOI{XXXXXXX.XXXXXXX}

\acmConference[Conference acronym 'XX]{Make sure to enter the correct
  conference title from your rights confirmation emai}{June 03--05,
  2018}{Woodstock, NY}
\acmPrice{15.00}
\acmISBN{978-1-4503-XXXX-X/18/06}

\begin{document}

\title{PORTAL: Portal Widget for Remote Target Acquisition and Control in Immersive Virtual Environments}

% \author{Submission ID: 9886}
\author{Dongyun Han}
\email{dongyun.han@usu.edu}
\affiliation{
  \institution{Utah State University}
  \city{Logan}
  \country{USA}}

\author{Donghoon Kim}
\email{donghoon.kim@usu.edu}
\affiliation{%
  \institution{Utah State University}
  \city{Logan}
  \country{USA}}

\author{Isaac Cho}
\email{isaac.cho@usu.edu}
\affiliation{%
  \institution{Utah State University}
  \city{Logan}
  \country{USA}}

\renewcommand{\shortauthors}{Han et al.}

\begin{abstract}
This paper introduces PORTAL (\textbf{PO}rtal widget for \textbf{R}emote \textbf{T}arget \textbf{A}cquisition and contro\textbf{L}) that allows the user to interact with out-of-reach objects in a virtual environment. We \revision{describe} the PORTAL interaction technique for placing a portal widget and interacting with target objects through the portal. We \revision{conduct} two formal user studies to evaluate PORTAL for selection and manipulation functionalities. The results show PORTAL supports participants to interact with remote objects successfully and precisely. Following that, we discuss its potential and limitations, and future works.
\end{abstract}


% \ccsdesc[500]{Human-centered computing~Empirical studies in HCI}


\begin{teaserfigure}
  \centering
  \includegraphics[width=0.9\textwidth]{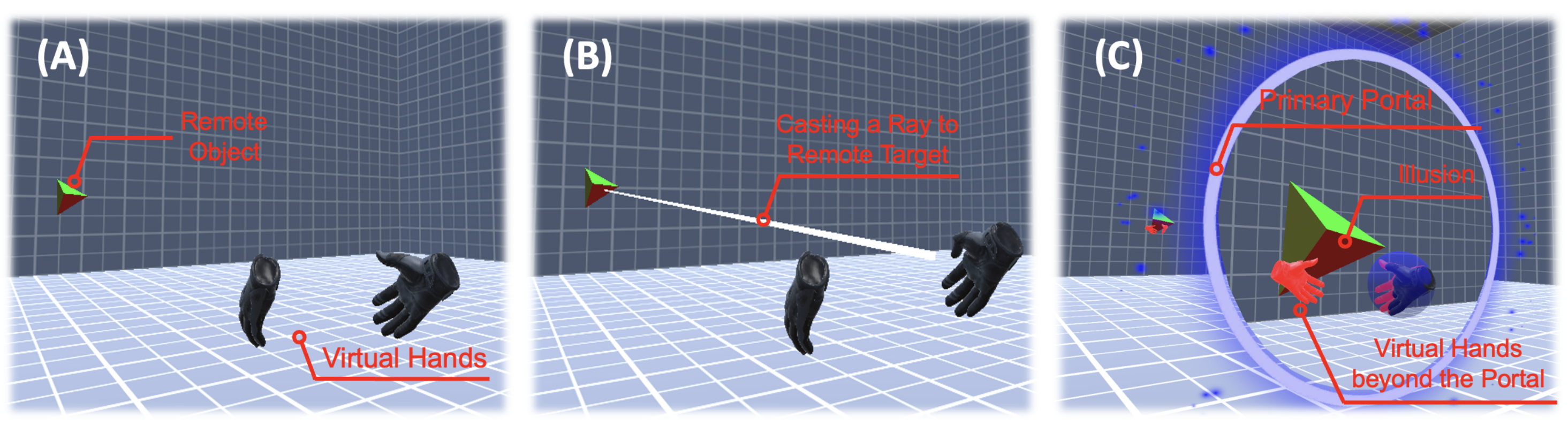}
  \caption{PORTAL is designed to leverage the secondary view interaction to allow the user to directly select and manipulate remote objects using simple virtual hands. (B) PORTAL uses ray-casting as its selection tool. (C) Once the user creates PORTAL by targeting a remote object, the user can see and interact with the remote object at a within-reach distance through PORTAL.}
  \label{portal_interaction}
\end{teaserfigure}

\maketitle

\section{Introduction}
% The 3D user interactions such as selection and manipulation in Immersive Virtual Environments (IVEs) have been studied and advanced by mimicking the way they operate in reality. 
% 3D user interactions such as selection and manipulation in Immersive Virtual Environments (IVEs) have been studied and advanced by mimicking the way they operate in reality. 
% However, interacting with out-of-reach objects in IVEs is challenging~\cite{laviola20173d}. 
3D user interactions for out-of-reach objects in Immersive Virtual Environments (IVEs) \revision{are} challenging~\cite{laviola20173d}. 
% Many techniques, including direct Hand-centered Object Manipulation Extending Ray-Casting (HOMER)\cite{bowman1997evaluation}, Worlds-in-Miniature (WIM)~\cite{stoakley1995virtual}, Go-Go~\cite{poupyrevgo}, and Linear Offset~\cite{li2015evaluation, li2018evaluation}, have been studied for remote object interactions~\cite{argelaguet2013survey, mendes2019survey}. 
Many techniques including direct HOMER~\cite{bowman1997evaluation}, Worlds-in-Miniature (WIM)~\cite{stoakley1995virtual}, Go-Go~\cite{poupyrevgo}, and Linear Offset~\cite{li2015evaluation, li2018evaluation}, have been studied for remote object interactions. However, they have limitations in terms of fine-grained object controls and precise depth perception for distant objects. % that are far away from the user.

To provide better depth perception for distant objects or locations, some techniques introduce a secondary view that is impossible to see through the user's first point of view in IVEs. 
For example, Kiyokawa and Takemura~\cite{kiyokawa2005tunnel} proposed Tunnel Window allowing the user to select and manipulate the remote objects through the window. 
% Recently, Li et al.~\cite{li2021vmirror} introduced vMirror supporting the user to select occluded objects by locating a virtual mirror in an IVE to show a different angle of the user's view. 
More recently, Li et al.~\cite{li2021vmirror} introduced vMirror supporting the user to select occluded objects by locating a virtual mirror to show a different angle of the user's view.
% Both approaches adopt ray-casting to select or manipulate the objects with the proposed technique. 
Both adopt ray-casting to select or manipulate objects with the proposed technique.
However, the usability of these techniques for \revision{distant} object selection and manipulation tasks is questionable because previous research~\cite{mine1997moving, li2015evaluation, paljic2002study} shows that direct selection with no offset outperforms varied distant object interaction techniques in terms of task completion time. 

% This paper introduces PORTAL, an interactive widget allowing the user to directly interact with remote objects using the simple hand metaphor by taking advantage of the secondary view interaction in IVEs.
This paper introduces PORTAL, allowing the user to directly interact with remote objects using the simple hand metaphor by taking advantage of the secondary view interaction. It presents an illusion that the remote objects are within-arm reach position and allows the user to interact with them by putting the hands into the portal. 
% We conducted two formal user studies to evaluate PORTAL. 
We conducted two formal user studies. Study 1 evaluates its usability and efficiency in selecting and manipulating out-of-reach objects by comparing PORTAL with existing techniques. In Study 2, PORTAL is compared with the direct interaction after teleporting to distant objects. The results show that PORTAL outperforms the direct HOMER and Linear Offset techniques and have a similar performance with Teleportation. Overall, the participants reported that they prefer PORTAL most as well as it is easy and enjoyable to use. The contributions of this paper are following:
\vspace{-0.1cm}
\begin{enumerate}[noitemsep]
% \vspace{-0.2cm}
% [topsep=0pt,itemsep=-1ex,partopsep=1ex,parsep=1ex]
% \item We introduce a novel interactive widget, PORTAL, to allow the user to directly interact with remote objects in IVEs. It is designed to support the user in selecting and manipulating distant objects with no offset and giving him or her better depth perception of remote locations.
% \item We introduce a novel interactive widget, PORTAL. It is designed to support the user in selecting and manipulating distant objects with no offset and giving him or her better depth perception of remote locations.
\item We introduce PORTAL, designed for the user to select and manipulate distant objects with no offset and a better depth perception of remote locations.
% \item We report our findings from two formal user studies to evaluate the efficiency and usability of PORTAL on the remote target selection and manipulation tasks. Study 1 compares PORTAL with two existing remote object interaction techniques: direct HOMER and Linear Offset. Study 2 compares PORTAL with direct interaction after teleporting to a remote position.
\item We conduct two user studies and report our findings to evaluate the efficiency and usability of PORTAL. 
% Study 1 compares PORTAL with two existing remote object interaction techniques: direct HOMER and Linear Offset. Study 2 compares PORTAL with Teleportation.
\item We provide a thorough discussion on our findings, the potential and limitations of PORTAL.  
\end{enumerate}

% This paper is organized as follows. In section 2, we review related works on 3D interaction techniques allowing remote interactions as well as the see-through interactions. We then describe PORTAL in section 3 along with the direct HOMER and Linear Offset techniques. In section 4 and 5, we present our experimental designs, and the results of the two user studies. 
% % In section 4, we present our experimental design, and the results of the user study. 
% We discuss the results, potential directions, and future works in section \ref{section_discussion}. We conclude our paper in section \ref{section_conclusion}.

% \vspace{-0.3cm}

\section{Related Works}

% \subsection{Remote Object Selection and Manipulation in IVE}
\subsection{Remote Object Interaction in IVE}
\label{remote_object_interaction_techniques}  
% Interacting with 3D objects in IVEs includes selecting, positioning, and rotating objects. 
% As these interactions frequently occur in our daily lives, there are a variety of scenarios to consider for designing 3D interaction techniques in Virtual Reality (VR) such as distances to objects, object sizes, density around objects, occlusion, etc~\cite{laviola20173d}.
% % These variables affect the technique performances. Examples are selection time and the total number of clicks until the user successfully selects a target object.
% \HDY{
% These variables affect user performances like interaction time and the total number of clicks required to reposition an object successfully.
% }

\begin{figure}[t]
\centering
\includegraphics[width=0.4\textwidth]{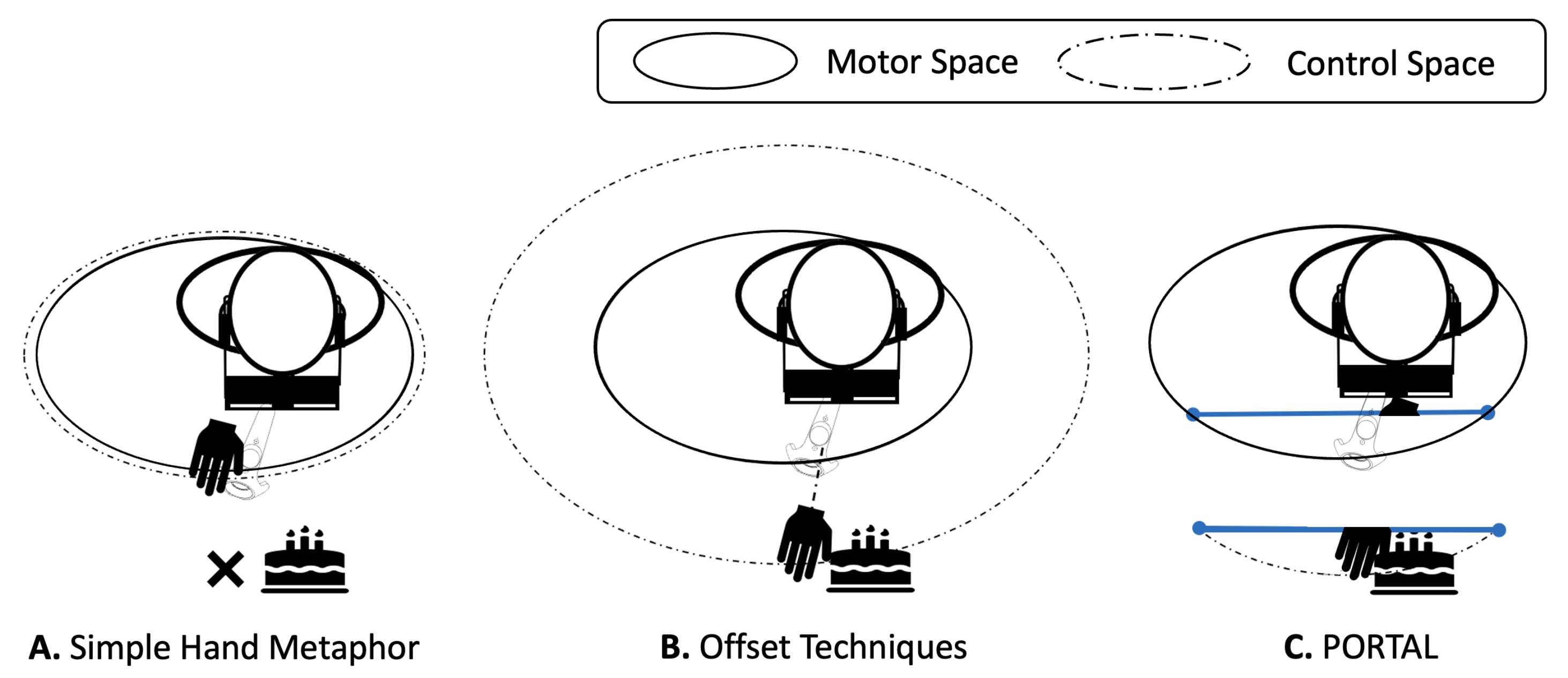}
\caption{(A) Simple hand metaphor can not reach remote objects because they are located out of reach. 
% B. The offset techniques such as direct HOMER, Go-Go, and Linear Offset allow the user to interact with the remote objects by extending the control space by multiplying a scale factor to the motor space.
% B. Offset techniques allow the user to interact with the remote objects by extending the control space by multiplying the motor space by a scale factor.
(B) Offset techniques extends the control space by multiplying the motor space by a scale factor. (C) PORTAL relocates a portion of the control space near the remote objects.}
\label{control_motor_space}   
\vspace{-0.4cm}
\end{figure}

To interact with objects in IVEs, the control and motor spaces should be considered~\cite{laviola20173d, argelaguet2013survey, mendes2019survey}. 
The control space refers to the spatial range in which the user is affordable to control objects, while the motor space is the physical space available for the user to operate the objects (i.e., arm-reach distance). 
% When the control space corresponds to the motor space (Figure~\ref{control_motor_space} A), the user cannot interact with a remote 3D object as it is positioned outside of the control space. However, the user can interact with the object when he or she moves to the remote position~\cite{bozgeyikli2016point, bolte2011jumper}. Teleportation~\cite{bowman1997travel} is a well known navigation technique in IVEs, allowing the user to move to the distant locations.
When the control space corresponds to the motor space (Fig.~\ref{control_motor_space} A), the user cannot interact with an out-of-reach object. 
% To interact with it, the user should move first to the remote position~\cite{bozgeyikli2016point, bolte2011jumper}. Teleportation~\cite{bowman1997travel} is a common navigation technique, allowing the user to move to the remote locations.
To interact with it, the user has to move to the remote position first ~\cite{bozgeyikli2016point, bolte2011jumper} using a navigation technique (e.g.,  Teleportation~\cite{bowman1997travel}.)
To avoid this, some interaction techniques provide a bigger control space than the user's arm-reach to select out-of-reach objects. An example is ray-casting, a widely adopted selection technique in many commercial VR and AR devices~\cite{ViveProducts, OculusProducts, HololensProducts}. 
It shoots a ray from a controller or the user's head to select a remote object that the ray hits. Another example is an image-plane pointing technique, also known as Sticky Finger~\cite{pierce1997image}. It allows the user to select a distant object that is occluded by using one of the index fingers. However, they provide limited support for object manipulation.

% There are two categorical approaches for reaching the user's control space to remote objects to fully support 6-DOF manipulation (xyz + yaw, pitch, roll) in IVEs. 
There are two categorical approaches to fully support remote 6 degree-of-freedom (DOF) object manipulation (xyz + yaw, pitch, roll) in IVEs. 
% The first approach is expanding the control space by multiplying the motor space by a scale factor or an offset that impacts on the control-display (CD) ratio.
The first approach is offset techniques (Fig.~\ref{control_motor_space} B), expanding the control space by multiplying the motor space by a scale factor that impacts on the control-display (CD) ratio.
The CD ratio determines how the input device's movements ($\Delta$x) are mapped to the virtual cursor's movement ($\Delta$X). It is defined as ($\Delta$x) / ($\Delta$X). 
% For example, if the CD ratio is 1 ($\Delta$x = $\Delta$X), the movement of the input device is equivalent to the movement of the virtual cursor (i.e., simple virtual hand interaction), and the control space is equivalent to the motor space (Figure~\ref{control_motor_space} A). 
For example, if the CD ratio is 1, the control space is equivalent to the motor space (i.e., simple virtual hand as Fig.~\ref{control_motor_space} A).
% Otherwise, when the CD ratio is smaller than 1 ($\Delta$x \textless $\Delta$X), the user can reach remote objects as the control space is expanded by the 1 / CD ratio, called offset techniques (Figure~\ref{control_motor_space} B). 
If the it is smaller than 1 ($\Delta$x \textless $\Delta$X), the user can reach remote objects as the control space is expanded by 1 / CD ratio. 
% Bowman et al.~\cite{bowman1997evaluation} proposed direct HOMER, and Poupyrev et al.~\cite{poupyrevgo} introduced Go-Go to support the user's interaction with remote objects in three-axis transitions and rotations, having a total of 6-DOF. 
Examples of this offset technique are Direct HOMER ~\cite{bowman1997evaluation},Go-Go ~\cite{poupyrevgo}, Linear Offset ~\cite{li2015evaluation, li2018evaluation}, Voodoo Dolls~\cite{pierce1999voodoo}, and World in Miniature (WIM)~\cite{stoakley1995virtual}.
%Bowman et al.~\cite{bowman1997evaluation} proposed direct HOMER, and Poupyrev et al.~\cite{poupyrevgo} introduced Go-Go.
%Other examples are the Voodoo Dolls~\cite{pierce1999voodoo} and WIM~\cite{stoakley1995virtual} techniques. Bluff~\cite{bluff2019don} recently introduced recursive interactions between real-life and miniature VR spaces which are similar to WIM.
%In addition, Li et al.~\cite{li2015evaluation, li2018evaluation} introduced the Linear Offset technique, which is a variation of Go-Go. 
These techniques have two limitations.  
% The first limitation is that precise control of out-of-reach objects is difficult to achieve due to the CD ratios. The second restriction, which applies to Voodoo Dolls and WIM particular, is that one hand must be tied to the widgets. 
1) precise control of out-of-reach objects is difficult due to the sensitive CD ratios. 2) for some techniques, one hand must be tied to the widgets (e.g., Voodoo Dolls and WIM). 

The second approach is the clutching mechanism~\cite{argelaguet2013survey} that relocates the control space to nearby target objects (Fig.~\ref{control_motor_space} C). It keeps 1 CD ratio for the user to interact with out-of-reach objects directly within a limited space. 
% In this paper, we adopt the clutching mechanism for PORTAL to enable the user to precisely interact with remote objects while keeping both hands free, overcoming the restrictions of the aforementioned techniques.
PORTAL adopts the clutching mechanism %, and it keeps the 1 CD ratio for the user to interact with out-of-reach objects directly within a limited space. 
to allow the user to precisely interact with remote objects while keeping both hands free in order to overcome the limitations of the aforementioned techniques.

% \subsection{See-through Interactions in IVE}
\subsection{Secondary View Interaction in IVE}
% See-through interactions provide a secondary view that is impossible to see through the user's primary view. 
% The primary view is the view that the user sees based on his or her head position in IVEs.
%The primary view is what the user sees based on his or her head position in IVEs. 
% \HDY{See-through interactions provide a secondary view that cannot be seen through the primary view, which the user sees based on his or her head position in IVEs.} 
% It helps the user identifies the virtual body or controllers as well as the surroundings. 
The secondary view is an additional view that displays a different perspective than the user's primary view to provide additional interactions for the user. The Magic Lens technique \cite{bier1993toolglass, viega19963d}, for example, reveals hidden information when it is overlaid on part of the primary view. 
% One applicable scenario for Magic Lens is the medical training~\cite{brown2006magic, schmalstieg1999sewing}, and the user can explore anatomy such as bones and organs when it is overlaid on a virtual human body. 
Its applicable scenario is the medical training~\cite{brown2006magic, schmalstieg1999sewing} to allow the user to explore anatomy (e.g., bones and organs) by overlaying the lens on a virtual human body.

Various types of secondary view interaction have been introduced. Photoportals~\cite{kunert2014photoportals} promotes telepresence between remote users in a virtual environment projected on 3D powerwalls 
%. It provides a method for remote users 
to take photos and videos together. Nam et al.~\cite{nam2019worlds} introduce Worlds-in-Wedges for visual comparison by rendering different VR scenes in multiple widgets. 

Secondary views are also studied for 3D object. Li et al.~\cite{li2021vmirror} proposed vMirror, an interactive widget leveraging reflection of mirrors to select remote and occluded objects. But it lacks object manipulation. Martin et al.~\cite{martin2020gain} also adopted a mirror metaphor providing a new perspective around an object for accurate object alignment. Stoev and Schmalstieg~\cite{stoev2002application} introduced Through-The-Lens (TTL) to allow the user to interact with remote objects through TTL with the limited DOF. 
% Kiyokawa and Takemura~\cite{kiyokawa2005tunnel} and Hirose et al.~\cite{hirose2006interactive} proposed Tunnel Window. 
% The user can dynamically open and close the window that supports the user to select and manipulate remote objects using ray-casting.
% It supports the user to select and manipulate remote objects using ray-casting, and the user can dynamically open and close the window. % In addition, users can bring the object over the tunnel window to them. 
To our best knowledge, only Tunnel Window~\cite{kiyokawa2005tunnel, hirose2006interactive} fully supports remote 6-DOF manipulation. It uses ray-casting as its selection tool. Compared to Tunnel Window, PORTAL uses a simple virtual hand metaphor allowing the user to interact with the target directly~\cite{ware1988using} through the secondary view. In addition we conducted two user studies to evaluate its efficiency and usability on the remote target selection and manipulation tasks while no evaluation was provided for Tunnel Window.

% % differences
% PORTAL utilizes the see-through interaction to support direct interaction with remote objects. 
% Compared to the previous works, PORTAL uses a simple virtual hand metaphor supporting 6-DOF interaction~\cite{ware1988using}. 
% Once a portal widget is opened targeting a remote object, it gives an illusion that the target object is relocated in a within-arm reach position. It allows the user to interact with the target directly without having to scale their controller movement (CD ratio = 1) through PORTAL.

% \section{Remote Interaction Techniques in IVE}
% \section{PORTAL: a widget for remote interaction}
\section{PORTAL: {\small a remote interaction widget}}
\label{distant_interaction_techniques}

\begin{figure}[t]
\centering
\includegraphics[width=.45\textwidth]{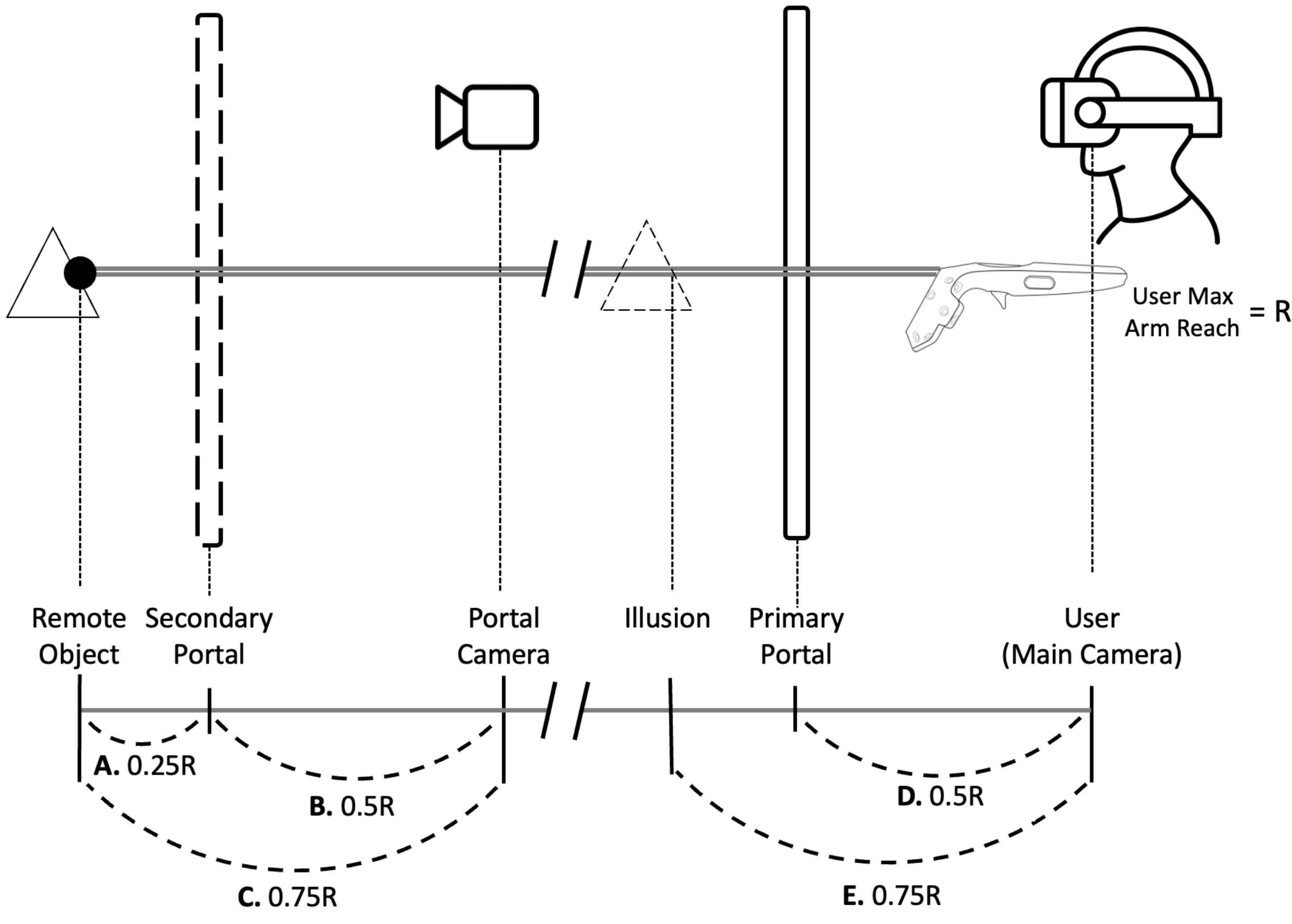}
% \caption{\HDY{PORTAL contains the primary portal and the secondary portal.} The portal camera is a child object in the secondary portal, and its view is rendered on the primary portal. Through PORTAL, the user perceive remote objects are in within-reach positions.} 
\caption{PORTAL contains the primary and the secondary portals. The portal camera in the secondary portal renders its view on the primary portal. Through PORTAL, the user perceive remote objects are in within-reach positions.} 
\label{portal_mechanism}   
 \vspace{-0.5cm}
\end{figure}

% PORTAL is an interactive widget leveraging a spatial tunnel or virtual portal metaphor for directly interacting with remote 3D  objects. 
PORTAL is an interactive widget leveraging a secondary view. 
% PORTAL is comprised of the primary portal and the secondary portal as shown in Figure~\ref{portal_mechanism}.
It is comprised of the primary portal and the secondary portal as shown in Fig.~\ref{portal_mechanism}.
The portals work as a spatial tunnel to connect different locations in IVEs to gives the user an impression that remote objects are right in front of him or her. The source codes and data are available at our GitHub repository
\footnote{\revision{\url{https://github.com/VIZ-US/PORTAL}}}.

% By putting hands into the primary portal, the user can take advantage of the simple hand metaphor that directly selects and manipulates the remote objects as if they were located within the user's reach.
% By putting hands into the primary portal, the user can directly select and manipulate the remote objects as if they were located within the user's reach.

\subsection{Design Considerations}
We design PORTAL with two considerations: 
1) the user's control space should reach remote objects and support direct selection and manipulation of them; 2) it should provide precise depth perception to the user to facilitate the fine-grained controls to the remote objects. We satisfy the first design consideration by adopting the clutching mechanism~\cite{argelaguet2013survey}. It relocates a portion of the user's control space to nearby target objects (Fig.~\ref{control_motor_space} C) and results in the user reaching out-of-reach objects in IVEs. The boundary of this control space is decided by the primary portal. When the user's virtual hands go beyond the primary portal, their copies are located at the distant position as shown in  Fig.~\ref{portal_interaction} C which follow the user's hand movement. The second consideration is achieved by projecting the secondary view from the portal camera on the primary portal. 
% It results the user to have the impression that the distant objects and their surroundings are right in front of him or her with precise depth perception.
Finally, the user gets the impression that the distant objects and their surroundings are right in front of him or her with precise depth perception.

\subsection{PORTAL Placement}
\begin{figure}[t]
\centering
\includegraphics[width=0.3\textwidth]{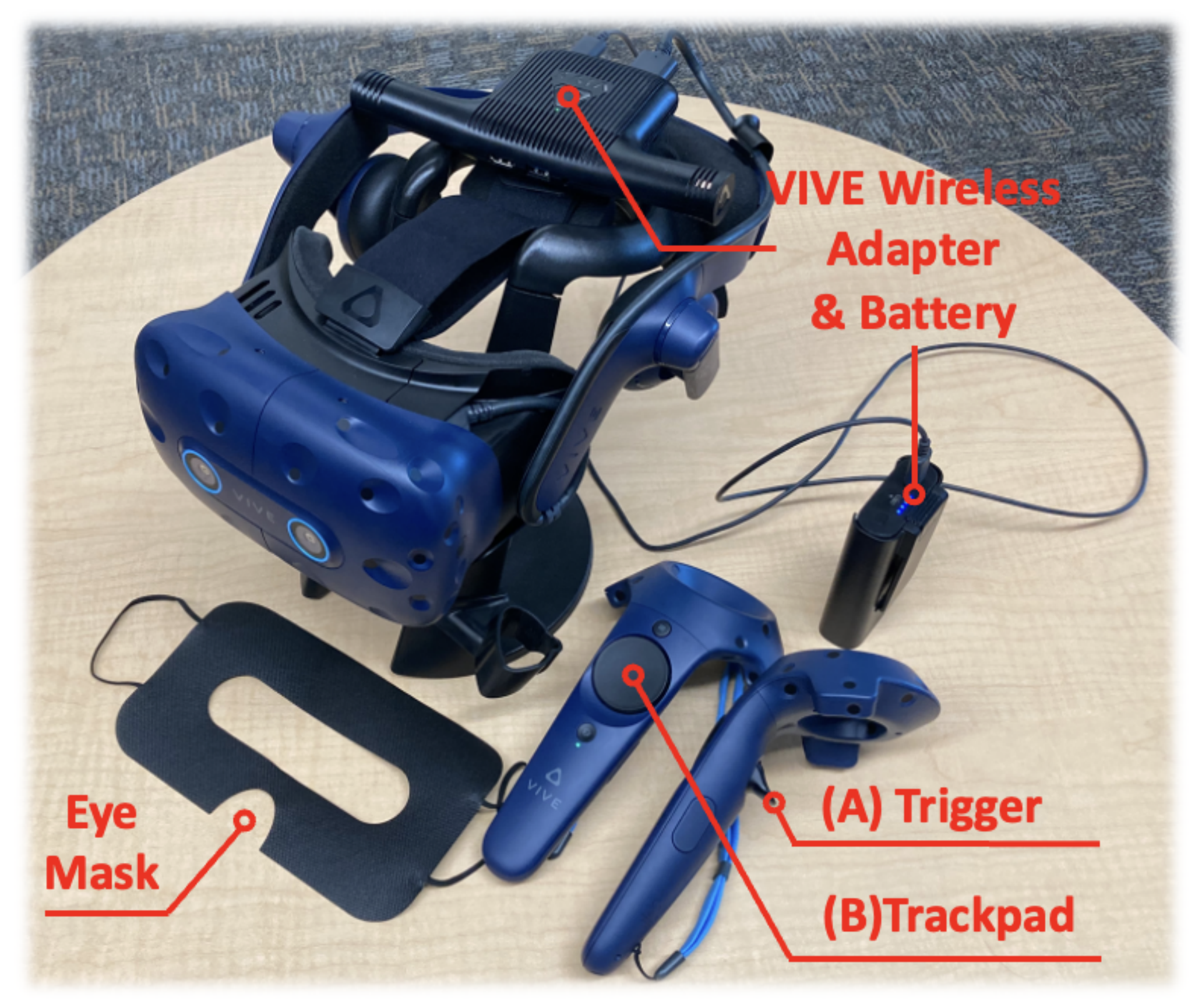}
% \caption{Study Apparatus. The Vive Pro Eye and two controllers are used. The Vive Pro Eye is connected to the Vive Wireless Adopter. Participants are asked to wear the eye mask to follow the COVID-19 procedures.}
\caption{Vive Pro Eye and its controllers are used. HMD is connected to Vive Wireless Adopter. Participants are asked to wear the eye mask to follow the COVID-19 procedures.}
\label{vr_devices}   
\vspace{-0.5cm}
\end{figure}

To place PORTAL, the user must select a remote target object by ray-casting. %  that is a target to interact with. PORTAL uses ray-casting as its selection tool. 
When the user presses the trackpad on the controller (Fig.~\ref{vr_devices} B), a ray emerges from the virtual hand as shown in Fig.~\ref{portal_interaction} B. 
When the user clicks the trigger button (Fig.~\ref{vr_devices} A) while the ray is pointing to the remote target, the primary and secondary portals are created. The primary portal is located in front of the user with the distance in the length of the user's arm reach (\textit{R}) \texttimes 0.5 in the ray’s direction 
(Fig.~\ref{portal_mechanism} D). 
% The primary portal is a circular shape, and its radius is 0.6m (Fig.~\ref{portal_interaction} C).
It is a circular shape, and its radius is 0.6m. The secondary portal is created in front of the remote target by \textit{R} \texttimes  0.25 in the opposite direction of the ray (Fig.~\ref{portal_mechanism} A). Their orientations are perpendicular to the ray. The secondary portal is invisible to the user. 

The view from the secondary portal is rendered on the primary portal. The portal camera is positioned in front of the target object by \textit{R} \texttimes 0.75 to the user direction (Fig.~\ref{portal_mechanism} C). Once the portal camera is placed in the position, it begins rendering its view on the primary portal by taking the following steps. We first find the projection matrices for the left- and right-views of the main VR camera in the scene. We apply the projection matrices to the portal camera and find the left- and right-eye projections which the portal camera sees. Each projection is mapped to the left or right texture on the primary portal object via Unity Shader. The distance between the user's head and the illusion of the target object through the primary portal (Fig.~\ref{portal_mechanism} E) is identical to the distance between the portal camera and the target object (Fig.~\ref{portal_mechanism} C). The portal camera moves following the user's head movement. It enables the user to see the target object and its surroundings with precise depth perception through the primary portal.

\subsection{PORTAL Interaction}
\label{subsection_portal_interaction}

% The user can interact with the remote target object by putting the virtual hands into the primary portal and grabbing it. 
The user can interact with a remote object by putting the virtual hands into the primary portal and grabbing the object like the one in front of the user.  
% A part of the hands that entered the primary portal is colored in red (Figure\ref{portal_interaction} C). 
A part of the hands that entered PORTAL is colored in red (Fig.\ref{portal_interaction} C). The red hands work the same as the simple virtual hands do. 
% The red hands in the portal work the same as the simple virtual hands do. 
% Within PORTAL, the user can directly select and manipulate the remote object as if it was in front of the user. 
Through PORTAL, the user can bring the object to the user's side and send an object to the remote side.

The user can reposition PORTAL like other 3D objects in IVEs. 
When the virtual hand is halfway through the primary portal, a blue spherical marker appears on the hand (Fig.\ref{portal_interaction} C), and it allows the user to grab the primary portal and relocate and rotate it. The primary portal's movement is linked to that of the secondary portal. As the portal camera belongs to the secondary portal object, when the user moves the primary portal, it changes the portal camera's position and orientation accordingly.
As a result, the projected view on the primary portal is changed. This interaction allows the user to navigate the remote location around the target object.

% The user can close PORTAL and update the view on the primary portal by changing the target. 
The user can close PORTAL and update the view on the primary portal. The same controller operation for creating PORTAL is used. The user can close PORTAL by performing the operation on the out of PORTAL. Conversely, when the user performs the operation targeting a new object within PORTAL, the position of the secondary portal is only changed. As a result, the view on the primary portal is updated, and it displays the new target.

% \section{ Study 1: {\small PORTAL vs. Remote Interaction Techniques}}
\section{ Study 1: {\large PORTAL vs. Existing Techniques}}

% In this section, we describe \HDY{our} Study 1 design and procedures.
% \HDY{This section describes \HDY{our} Study 1 design and procedures.} 
Study 1 (IRB \#12168) includes two tasks to evaluate PORTAL compared to the remote interaction techniques including direct HOMER (HOMER) and Linear Offset (LO). Please note that the objective of Study 1 is to evaluate the usability and efficiency of PORTAL on remote object selection and manipulation tasks. 
% We focus on evaluating the usability and efficiency of PORTAL on selection and manipulation tasks. 
% So, the functions manipulating the primary portal and sending and receiving an object through the portal are deactivated in Study 1.
So, the functions relocating PORTAL and sending and receiving an object through PORTAL are deactivated in Study 1. Study 1 takes approximately one hour.

\subsection{\large Direct HOMER and Linear Offset Techniques} \label{other_techniques}

% In Section~\ref{remote_object_interaction_techniques}, we reviewed the multiple interaction techniques for remote objects in IVEs. 
In Section~\ref{remote_object_interaction_techniques}, we reviewed the remote object interaction techniques. Among them, we carefully selected techniques for Study 1 satisfying the following criteria: 
1) Techniques supporting fully 6-DOF manipulation using one hand. So techniques like Voodoo Dolls and WIM are excluded; 
% 2. Direct techniques to select and manipulate objects rather than indirect ways.
% 2) Techniques using the same device modality. We use the common VR controllers (Fig~\ref{vr_devices}). 
2) Techniques using the same device modality. As we use the Vive controllers (Fig~\ref{vr_devices}) 
% So, techniques using eyes or hand tracking are not considered; 
, eyes or hand tracking techniques are not considered; 
% 3) Techniques showing better results in terms of performance and user preference reported in the previous research. Among the offset techniques, we choose LO instead of Go-Go for Study 1 because Li et al.~\cite{li2015evaluation, li2018evaluation} showed that LO outperforms Go-Go in remote object selection and navigation. In the same manner, we remove the Sticky Finger technique as a selection tool from our list as it has been reported that it perform worse than ray-casting~\cite{bowman2001testbed, chow2008wii}. As a result, the techniques for Study 1 are PORTAL, HOMER and LO. 
3) Techniques showing better performance and user preference in the previous research. Among the offset techniques, we choose LO instead of Go-Go because Li et al.~\cite{li2015evaluation, li2018evaluation} showed that LO outperforms Go-Go in remote object selection and navigation. In the same manner, Sticky Finger is excluded from our list because it performs worse than ray-casting~\cite{bowman2001testbed, chow2008wii}. As a result, we use  PORTAL, HOMER and LO for Study 1. 

\textbf{Direct HOMER} is a hybrid technique of ray-casting and the virtual Hand metaphor. 
% It uses ray-casting as its selection tool, so the user sees a ray shot from one of the virtual hands in IVEs. 
It uses ray-casting as its selection tool. The user can select a remote object which the ray hits. When the user selects the remote object, it is not attached to the ray, but instead a virtual hand that shoots the ray moves to the target instantly. 
% It supports the 6-DOF manipulation. 
Once the user selects the target, an offset is applied to the virtual hand (CD ratio \textless 1) following the user's controller movement, while its rotation corresponds one-to-one. The offset is determined by the distance between the user and the controller over the distance between the user and the object at the moment of selection. The virtual hand returns to the user when the interaction is done.

\textbf{The Linear Offset technique} maps the offset to virtual cursors by the distance between the user and controllers, regardless of the threshold distance~\cite{li2015evaluation}. 
% It also supports 6-DOF manipulation.
Compared to direct HOMER, the offset ratio is fixed, so it is not varied by the distance from the user and target objects. In Study 1, LO is properly controlled for each user, allowing the virtual hands to reach the virtual room's walls from the center of the room when the user fully extends his or her arms.

\subsection{Apparatus}

% Study 1 is performed using a Vive Pro Eye with a Vive Wireless Adapter and two controllers (Figure~\ref{vr_devices}). The Vive Pro Eye headset has a \degree{110} Field-of-View, a resolution of 1440 \texttimes 1600 for each eye, and 90 Hz refresh rate. 
Study 1 is performed using a Vive Pro Eye with a Wireless Adapter and two Vive controllers (Fig.~\ref{vr_devices}). The headset has a 110\degree{} Field-of-View, a resolution of 1440 \texttimes 1600 for each eye, and a 90 Hz refresh rate. 
% The wireless adapter provides near-zero latency~\cite{VIVE_WIRELESS_ADAPTER} \footnote{https://www.vive.com/us/accessory/wireless-adapter/}. 
The wireless adapter provides near-zero latency~\cite{ViveWirelessAdaptor}. 
% The VR application for the two tasks is implemented in Unity 2019.4.31f1 and ran on a Window 10 desktop which has Intel Xeon W-2245 CPU (3.90GHz), 64GB RAM, and Nvidia GeForce RTX 3090 graphics card. 
The VR application for Study 1 is implemented in Unity 2019.4.31f1 and ran on a Window 10 desktop which has Intel Xeon W-2245 CPU (3.90GHz), 64GB RAM, and Nvidia GeForce RTX 3090 graphics card. 
% Object selection and manipulation in our VR application are accomplished by pressing the trigger button on the controllers and rotating the controllers afterward.
%In the application, a trigger button is used to select and manipulate an object. 
%on a controller and then moving it. 

\subsection{Participants}

A total of 21 participants (14 males and 7 females, 19 right handed and 2 left handed) were recruited from the university. 
% The average age of the participants is 21.2, ranging from 18 to 27.
Their average age is 21.2, ranging from 18 to 27. 
Fifteen of them were recruited through the e-mail recruitment process from the Computer Science department, while six were recruited through our university's student participants recruitment system (SONA) from Statistics, Accounting and Finance, and Physics departments. The participants were rewarded with the same value as stipulated in the university SONA policy (1 SONA point and a \$10 gift card for the participants from SONA and the e-mail recruitment processes respectively). %Nineteen participants are right-handed, and two participants are left-handed. 
We measured the participants' arm reach for PORTAL and LO. The average arm reach from their shoulder to the dominant hand is 61.43cm. 
All participants have 20/20 (or corrected 20/20) vision and have no impairments in the use of VR devices. Twenty participants reported that they have used VR devices before, and their average self-VR familiarity score is 3.9 out of 7.0.

\subsection{Procedures}
Upon the arrival, a participant signs the informed consent form. 
% Then the participant is briefed with the experimental procedures, and completes a pre-questionnaire asking a demographic questionnaire. 
Then the participant is briefed on the experimental procedures and asked to complete a demographic questionnaire. Next, the participant is introduced to PORTAL, HOMER, and LO, and their controller operations. The participant has a 5-10 minutes training session to be familiar with the techniques. It involves interacting with three 3D cubes in different locations in an IVE. Before wearing the headset, the participant is required to wear an eye mask (Fig.~\ref{vr_devices}). 
% Study 1 is approved from IRB (IRB \#12168).

Next, the participant is introduced to Task 1 and 2. The participant is asked to do the tasks as quickly and accurately as possible. The participant first conducts Task 1, followed by Task 2. 
% Before beginning a task, the participant wears the headset and tightens it around his or her head.
% When the VR application is run, the participant is asked to move to the center of the VR room (a red ``X" mark with an arrow on the floor), which is 20m \texttimes  20m \texttimes  10m (width \texttimes  length \texttimes  height) and colored in sky blue with white cross checks as shown in Figure~\ref{portal_interaction}. 
When the VR application is run, the participant is asked to move to the center of the VR room, where is a red ``X" mark with an arrow on the floor. Then, the participant can see a green box with a panel showing the technique to be used and the current degree of completion.
The participant can start the trial by clicking the green box. The box then disappears, and task objects appear. 
% After completing a task using one of three techniques, the participant takes the headset off and completes a survey evaluating the technique's effectiveness such as the levels of arm fatigue and self-success on a 7-point Likert scale ranging from 1 (Very Low) to 7 (Very High).
After completing a task using one of three techniques, the participant takes the headset off and completes a survey evaluating the technique including the levels of arm fatigue and self-success on a 7-point Likert scale ranging from 1 (Very Low) to 7 (Very High). 
After the survey, the participant puts on the headset again, moves to the center of the VR scene, and proceeds to new trials with the next technique.
When all the tasks are completed, the participant completes a post-questionnaire.
It asks the participant to rate the techniques based on how easy they are to select and dock the targets on a 7-point Likert scale ranging from 1 (Very Easy) to 7 (Very Difficult). 
%  The participant also ranks three techniques in order of their preferences, from 1 (the best) to 3 (the least).
The participant also ranks them in order of their preferences, from 1 (the best) to 3 (the least). After each participant finished Study 1, we sanitized all VR equipment and tables. We had at least 30 minutes intervals for the next participant to follow the COVID-19 procedures.

\begin{figure}[t]
\centering
\includegraphics[width=0.28\textwidth]{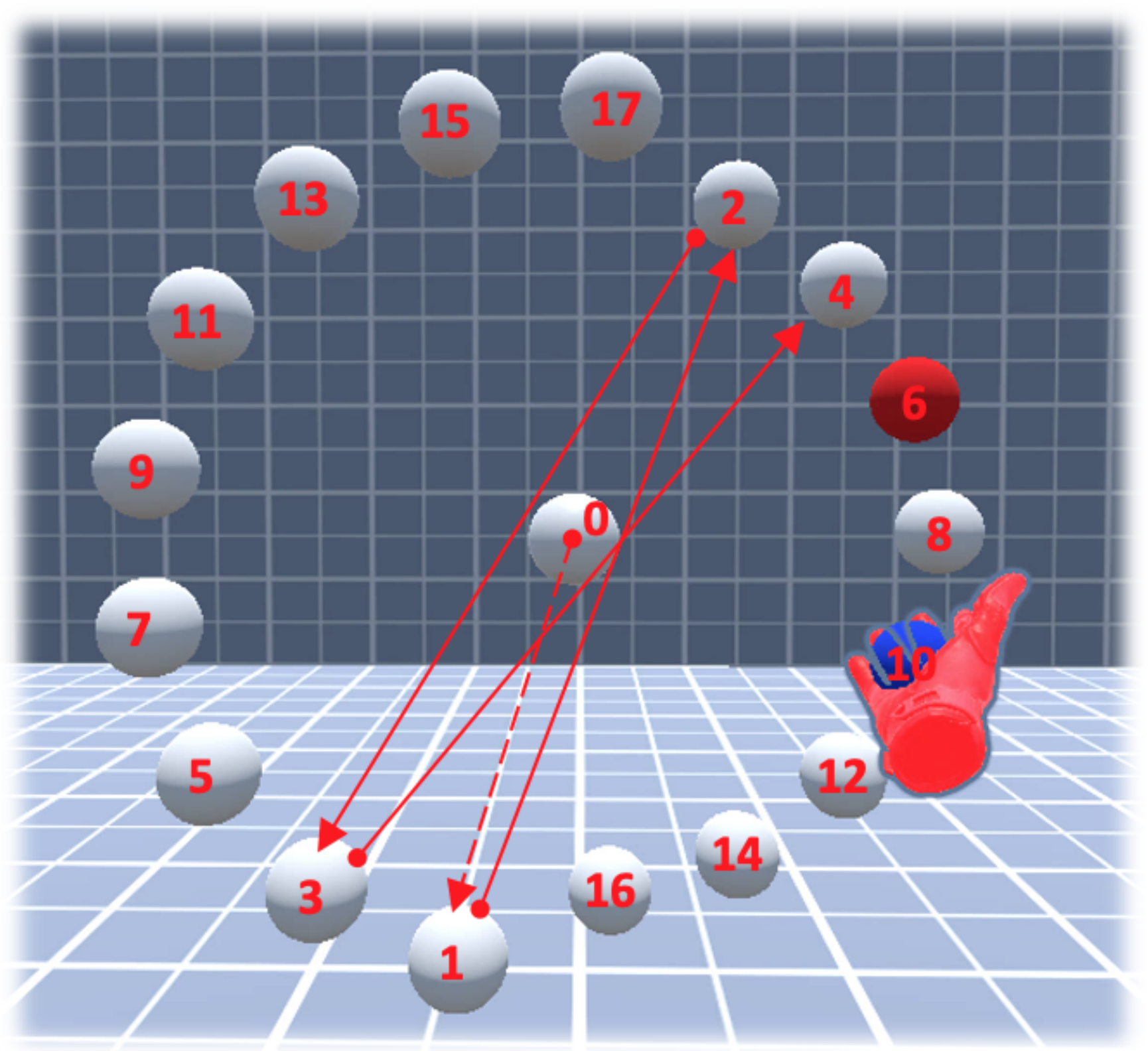}
% \caption{The selection task layout follows ISO 9241-9. Its diameter is 60 cm, and each target width is 7 cm, having the 3.26 ID. }
\caption{The task 1 layout follows ISO 9241-9. Its diameter is 60 cm, and each target width is 7 cm, having the 3.26 ID. }
\label{selection_task_layout}   
\vspace{-0.4cm}
\end{figure}

\subsection{Tasks}
\subsubsection{Task1}
% Task 1 is objective to evaluate the efficiency and usability of PORTAL in terms of the selection time and the number of clicks required to acquire targets within PORTAL. 
The objective of Task 1 is to evaluate PORTAL in terms of the selection time and the number of clicks required to acquire targets correctly. 
Task 1 is inspired from the Fitts' law test (ISO 9241-9), the multi-directional tapping task with a circular arrangement of targets. 
% The task object follows ISO 9241-9 with one additional sphere in the center.
% \HDY{It} follows ISO 9241-9 with one additional sphere in the center.
It consists of one sphere in the center and the 17 spheres arranged in a circular layout having a diameter of 60cm with varying depths. The arrows in Fig.~\ref{selection_task_layout} show the consecutive paths that the participants should follow to select the next targets clockwise. By targeting the center sphere, the participant can open PORTAL in front of the targets. The spheres have a radius of 7cm. The nine spheres in the left half are located with a depth of -7cm relative to the central sphere, while the eight spheres in the right half are marked with a depth of +7cm. We change the depth between subsequent targets following Teather and Stuerzlinger's work~\cite{teather2013pointing}. The spheres are displayed in one of three colors as Fig.~\ref{selection_task_layout}. The target sphere is colored in red, and the others are in white. When the virtual hand hovers over or the ray hits a sphere, it turns blue. 
% It provides the participant with a visual cue representing what object the participant is attempting to select. 
It serves as a visual cue to the participant as to which object he or she is attempting to select. When the participant selects a target correctly, the next target is highlighted in red, and the previously selected target is de-highlighted in white.   

Task 1 uses a within-subject design with two independent variables: Technique (PORTAL, HOMER, and LO) and Target Distance (3m, 6m, and 9m). The dependent variables are the selection time, error rate, and throughput. The selection time measures how long the participant takes to select the target after the previous target was chosen. The error rate refers to the number of clicks to successfully select the target. Throughput is the measurement that quantifies the human rate of information processing in bits per second (bps) that can be measured by the average value of index of difficulty (ID) over the average time of the selections. In Task 1, we only change the target distance, instead of changing ID by manipulating the size of spheres or circular layout. Please note that this study is not objective to model performance as a function of target sizes and circular layout. It allows us to evaluate three techniques in a manageable number of study conditions. The ID in Task 1 is 3.26, and it is kept in the target distances.

A set of trials consists of one center sphere selection (Sphere 0 in Fig.~\ref{selection_task_layout}) and sixteen target selections. 
% Logging starts when the participant selects the green cube at the center of our VR scene. 
Each participant completes 27 center sphere selections (3 target distances × 3 techniques × 3 trials × 1 selection) and 432 target selections (3 target distances × 3 techniques × 3 trials × 16 selections). The selection of Sphere 1 after clicking Sphere 0 is not counted because its movement (the dotted line in Fig.~\ref{selection_task_layout}) is almost half compared to the other selections. 
% The center sphere selection measures the time taken for the participant to reach the remote objects. 
The Sphere 0 selection is used to measure time taken for the participant to reach the remote objects. In the PORTAL case, it allows us to track how long it takes the participant to open PORTAL correctly. Sequences of the techniques and the target distances are balanced across all the participants using a Latin square design.

\subsubsection{Task 2: Docking}

\begin{figure}[t]
\centering
\includegraphics[width=0.3\textwidth]{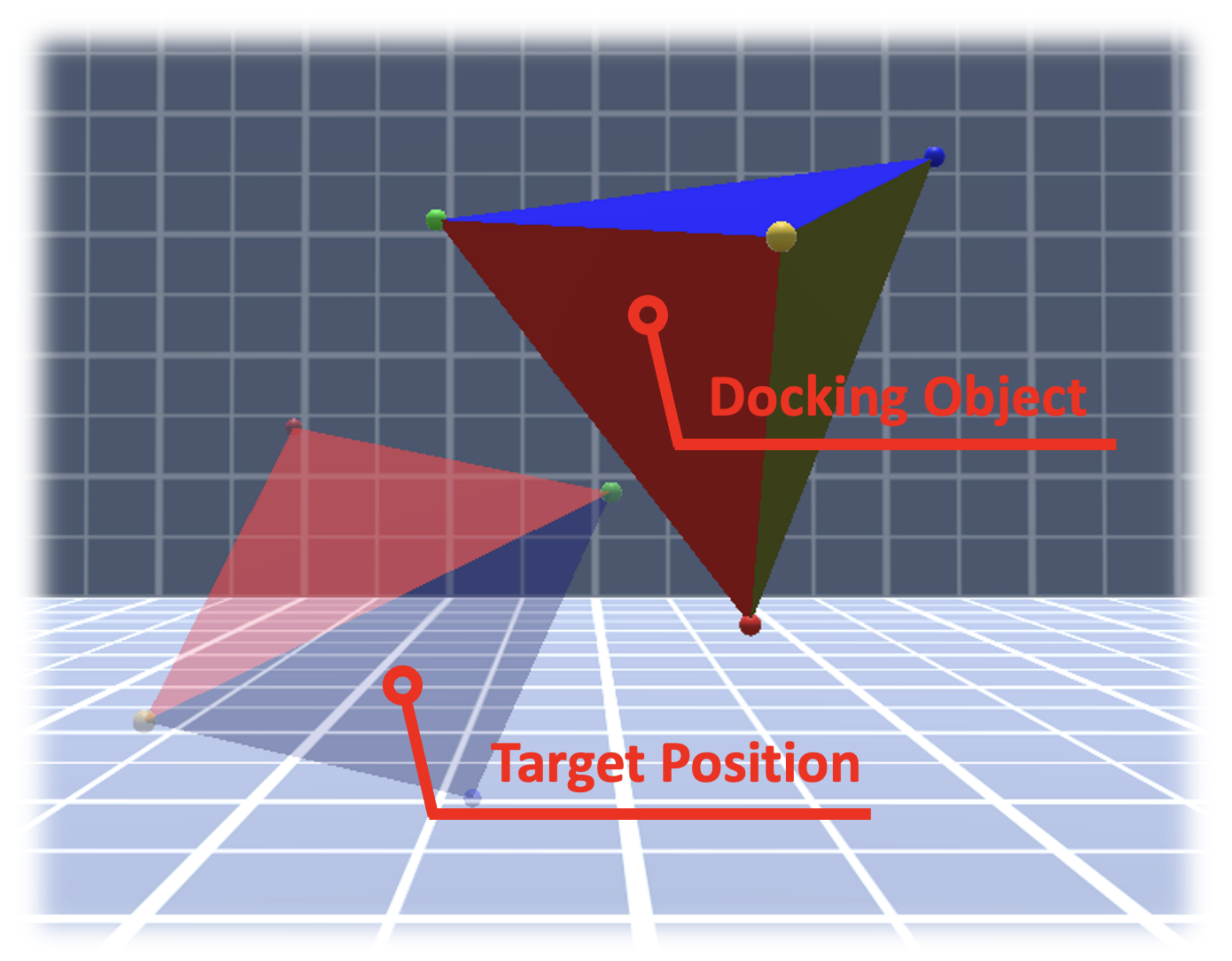}
% \caption{Task 2 consists of a docking object and its target position. The docking object and the target position are represented in the opaque and semi-transparent tetrahedron objects respectively. } 
\caption{Task 2 consists of a docking object and its target position which are represented in the opaque and semi-transparent tetrahedron objects respectively. } 
\label{docking_task_layout}   
\vspace{-0.3cm}
\end{figure}

We designed Task 2 based on previous research ~\cite{zhai1993human, li2015evaluation, froehlich2006globefish}. It consists of two tetrahedron objects as shown in  Fig.~\ref{docking_task_layout}. The opaque tetrahedron is an object which the participants are asked to reposition, and the semi-transparent tetrahedron represents the target position where the object should be docked. 
% They are identical except the opacity. 
They are regular tetrahedrons which on each side is 0.5m. They have a different color on each face and vertex sphere. The trials consist of repositioning the docking object to the target position considering the orientation and colors. 
% The participant is asked to dock the object as quickly and accurately as possible. 
The position and orientation of the target position are randomly determined across the trials within a 0.5m radius of the docking object. To complete a trial, the participant must align their corresponding vertices within a tolerance of 4.5cm which is decided through the pilot study. The docking object's outline turns green when it is positioned in the target position within the tolerance. 
% The participant can complete the trial and move on to the next trial when the outline turns green by clicking the controller's trigger on their minor hand.
The participant can move on to the next trial when the outline is green by clicking the controller's trigger on the minor hand.

Task 2 employs a within-subject design with two independent variables, Technique (PORTAL, HOMER, and LO) and Target Distance (3m, 6m, and 9m).
The dependent variables are the \revision{docking} time and the error distance. The \revision{docking} time measures how long it takes to move the docking object to the target position after the user grabbed it. The error distance is calculated by adding the distances between the vertices of two tetrahedrons. 
Each participant completes 27 trials in Task 2 (3 target distances × 3 techniques × 3 trials). Sequences of Technique and Target Distance are counter-balanced. 
% Each participant must complete 27 trials in Task 2 (3 target distances × 3 techniques × 3 trials). Sequences of Technique and Target Distance are counter-balanced. 

\subsubsection{Hypothesis}
The main hypothesis of Study 1 are:

\begin{enumerate}
% [label=H\arabic*. , wide=0.5em,  leftmargin=*, topsep=0pt, partopsep=0pt]
[label=H\arabic*. ,noitemsep]
\item The further away the target objects are, PORTAL is expected to have a faster target selection/docking completion time than HOMER and LO, once PORTAL is opened. This is because the user can directly select objects within PORTAL.
\item PORTAL is expected to have a lower error rate/distance than HOMER and LO with all target distances because PORTAL allows the user to interact with remote objects directly.
\item PORTAL is expected to perform the same regardless of target distance because the user may feel that the targets are located at the same within-reach proximity once the portal is opened.
\end{enumerate}

\subsection{Result}
We report the results analyzed with a two-way repeated measures Analysis of Variance (ANOVA) test at the 5\% significance level. 
The degrees of freedom are corrected using Greenhouse-Geisser correction to protect against violations of the sphericity assumption. 
The post-hoc test is performed using Tukey’s honestly significant difference (HSD) comparisons at the same 5\% significance level.
\subsubsection{Task1: Quantitative Result}
Please note that we analyze the target selections except the Sphere 0 and 1 selections in the following. We discussed the Sphere 0 and 1 selections in Section~\ref{sec_time_for_opening_portals}.

\begin{figure}[t]
\centering
\includegraphics[width=.5\textwidth]{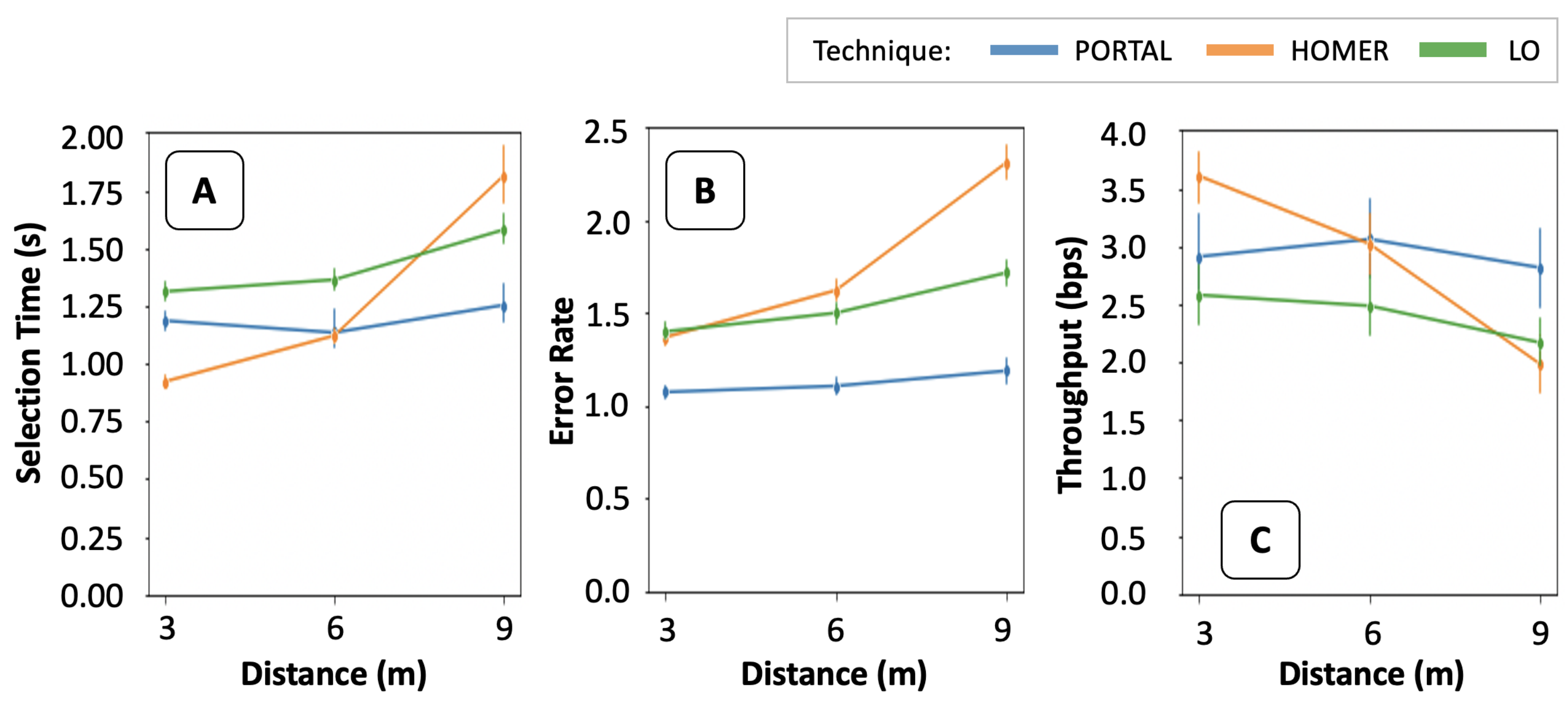}
% \caption{Task 1 Results: Interaction effects on the selection times, error clicks, and throughput with the error bars representing 95\% CI.} 
\caption{Task 1 Results. Interaction effects on the selection times, error clicks, and throughput with 95\% CI error bars.} 
\label{selection_task_results}   
\vspace{-0.3cm}
\end{figure}

\begin{table}[ht]
\caption{\label{table_selection_completionTime_result} \textbf{Task 1: Significant effects on Selection Time}}
\begin{tabular} {m{2.5cm} | m{1cm} | m{1cm}| m{1cm} | m{0.8cm}}
\toprule
    {\textbf{Factor}} & {\textbf{$DOF$}}  & {\textbf{$F$}} & {\textbf{$p$}} & {\textbf{$\eta$\textsubscript{$p$}\textsuperscript{$2$}}}\\ \midrule
    Technique\texttimes Distance  & {4, 64} & {9.344}  & {.002} & {.369} \\ 
    Technique  & {2, 32} &  {4.59}  & {.026} & {.0223} \\
    Distance & {2, 32}  & {21.6} & {\textless .001} & {.574} \\
    \bottomrule
\end{tabular}
\vspace{-0.3cm}
\end{table}

\textbf{Selection Time:} The statistical results on selection time are reported in Table~\ref{table_selection_completionTime_result}. 
% The result of a two-way repeated measures ANOVA shows a significant interaction effect on selection time between Technique and Target Distance ($p$=.002, Figure 6 A).  
The result shows a significant interaction effect on selection time between Technique and Target Distance ($p$=.002, Fig.\ref{selection_task_results} A). There is a simple effect of Technique with Target Distance at 3m ($p$\textless.001), 6m ($p$=.020), and 9m ($p$=.005). 
At 3m, HOMER (M=.910s) has a faster selection time than PORTAL (M=1.185s, $p$\textless.001), but no difference between PORTAL and LO (M=1.312s, $p$=.129). 
With Target Distance 6m, PORTAL (M=1.134s, $p$\textless.001) and HOMER (M=1.121s, $p$\textless.001) are significantly faster than LO but no statistical difference between HOMER and PORTAL ($p$\textgreater.900) is found.
%, while LO (M=1.363s) shows a slower selection time than . 
% When Target Distance is 9m, PORTAL (M=1.253s) selects the targets significantly faster than HOMER (M=1.814s, $p$\textless.001) and LO (M=1.579s, $p$\textless .001). 
When Target Distance is 9m, PORTAL (M=1.253s) is significantly faster than HOMER (M=1.814s, $p$\textless.001) and LO (M=1.579s, $p$\textless .001). LO is also faster than HOMER ($p$=.001). %in the 9m condition. 
% There is no simple effect of Target Distance with PORTAL ($p$=.523). 
PORTAL has no simple effect on Target Distance ($p$=.523). 

There is a main effect of Technique on selection time ($p$\textless .001). The pairwise comparison shows that overall PORTAL (M=1.194s) has a faster selection time than HOMER (M=1.261s, $p$=.041) and LO (M=1.41s, $p$\textless.001) and
% In addition, HOMER is faster than LO ($p$\textless.001). 
HOMER is faster than LO ($p$\textless.001).

\begin{table}[ht]
% \captionof{table}{Table Title} \label{tab:title} 
\caption{\label{table_selection_error_rate_result} 
\textbf{Task 1: Significant effects on Error Rate}}
% \vspace{-0.1cm}
\begin{tabular} {m{2.5cm} | m{1cm} | m{1cm}| m{1cm} | m{0.8cm}}
\toprule
    {\textbf{Factor}} & {\textbf{$DOF$}}  & {\textbf{$F$}} & {\textbf{$p$}} & {\textbf{$\eta$\textsubscript{$p$}\textsuperscript{$2$}}}\\ \midrule
    Technique\texttimes Distance  & {4, 64} & {17.927}  & {\textless .001} & {.528} \\ 
    Technique  & {2, 32} &  {22.53}  & {\textless.001} & {.585} \\
    Distance  & {2, 32}  & {19.259} & {\textless.001} & {.546} \\
    \bottomrule
\end{tabular}
\vspace{-0.3cm}
% \label{table:1}
\end{table}

\textbf{Error Rate:} 
Table~\ref{table_selection_error_rate_result} shows the statistical results on error rates. There is a significant interaction effect between Technique and Target Distance ($p$\textless.001). We found a simple effect of Technique with Target Distance 3m ($p$=.011), 6m ($p$=.001), and 9m ($p$\textless.001) ( Fig.\ref{selection_task_results} B). 
PORTAL (M\textsubscript{$3m$}=1.107, M\textsubscript{$6m$}=1.136, M\textsubscript{$9m$}=1.188) shows the significantly lower error rates than HOMER (M\textsubscript{$3m$}=1.37, $p$=.001, M\textsubscript{$6m$}=1.673, $p$\textless.001, and M\textsubscript{$9m$}=2.162, $p$\textless.001) and LO (M\textsubscript{$3m$}=1.4,$p$=.001, M\textsubscript{$6m$}=1.451, $p$\textless.001, and M\textsubscript{$9m$}=1.617, $p$\textless.001) at all the distances.
In addition, LO shows a significantly lower error rate than HOMER at 9m ($p$\textless.001) while there are no statistical differences between HOMER and LO at 3m and 6m. For PORTAL, no simple effect of Target Distance is found ($p$=.278).

\begin{table}[h]
% \captionof{table}{Table Title} \label{tab:title} 
\caption{\label{table_selection_throughput_result} \textbf{Task 1: Significant effects on Throughput}}
\begin{tabular} {m{2.5cm} | m{1cm} | m{1cm}| m{1cm} | m{0.8cm}}
\toprule
    {\textbf{Factor}} & {\textbf{$DOF$}}  & {\textbf{$F$}} & {\textbf{$p$}} & {\textbf{$\eta$\textsubscript{$p$}\textsuperscript{$2$}}}\\ \midrule
    Technique\texttimes Distance  & {4, 64} & {12.716}  & {\textless .001} & {.443} \\ 
    Technique  & {2, 32} &  {9.691}  & {.001} & {.377} \\
    Distance  & {2, 32}  & {37.086} & {\textless .001} & {.699} \\
    \bottomrule
\end{tabular}
\vspace{-0.3cm}
\end{table}

\textbf{Throughput:}
The statistical results on throughput is shown in Table~\ref{table_selection_throughput_result}. 
It shows a significant interaction effect on throughput between Technique and Target Distance ($p$\textless.001).
We find a simple effect of Technique with Target Distance at 3m ($p$\textless.001), 6m ($p$=.016), and 9m ($p$\textless.001) respectively as shown in Fig.\ref{selection_task_results} C. At 3m, HOMER (M=3.606 bit/s) has a higher throughput than PORTAL (M=2.910 bit/s, $p$=.019) and LO (M=2.579 bit/s, $p$\textless.001). 
At 6m, PORTAL (M=3.061 bit/s) and HOMER (M=3.019 bit/s) show the statistical differences to LO (M=2.485 bit/s, $p$\textsubscript{PORTAL}=.019, $p$\textsubscript{HOMER}=.031). When Target Distance is 9m, PORTAL (M=2.815 bit/s) has a higher throughput than HOMER (M=1.984 bit/s, $p$\textless.001) and LO (M=2.168 bit/s, $p$\textless .001). According to PORTAL, it shows the constant throughput by Target Distance ($p$=.517). 

A main effect of Technique is found ($p$\textless .001). The pairwise comparison shows that PORTAL (M=2.928 bit/s) and HOMER (M=2.87 bit/s) have a higher throughput than LO (M=2.41 bit/s,  $p$\textless .001), but no difference is disclosed between PORTAL and HOMER ($p$=.856). 

\subsubsection{Task 2: Quantitative Result}
 We report the statistical results of the \revision{docking} times and error distances.

\begin{figure}[t]
\centering
\includegraphics[width=0.38\textwidth]{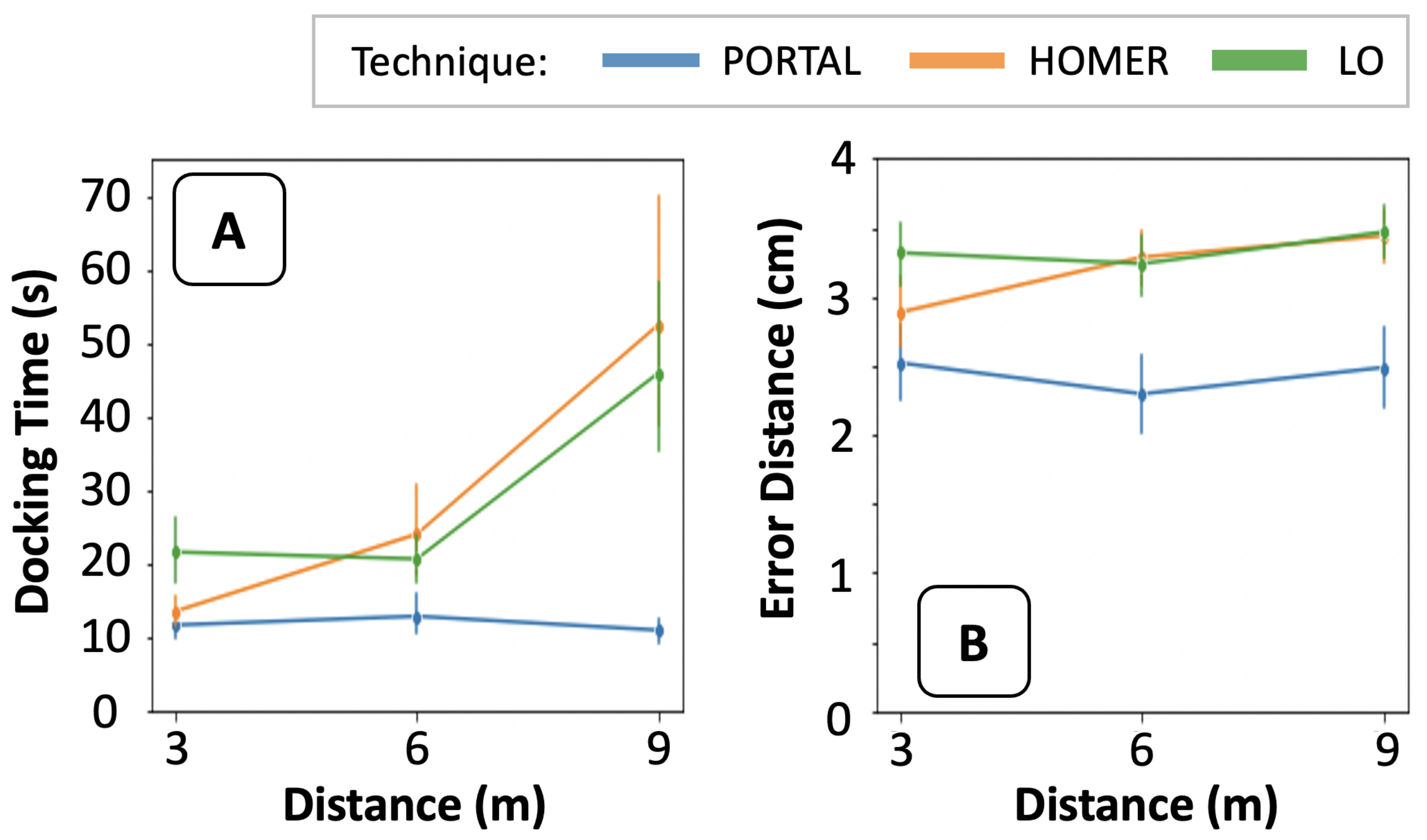}
\caption{Task 2 Results. Interaction effects on the \revision{docking} times and the error distance with 95\% CI error bars. } 
\label{docking_task_results}   
% \vspace{-0.3cm}
\end{figure}

\begin{table}[t]
% \captionof{table}{Table Title} \label{tab:title} 
\caption{\label{table_docking_completionTime_result} \textbf{Task 2: Significant effects on \revision{Docking} Time}}
\begin{tabular} {m{2.5cm} | m{1cm} | m{1cm}| m{1cm} | m{0.8cm}}
\toprule
    {\textbf{Factor}} & {\textbf{$DOF$}}  & {\textbf{$F$}} & {\textbf{$p$}} & {\textbf{$\eta$\textsubscript{$p$}\textsuperscript{$2$}}}\\ \midrule
     Technique\texttimes Distance  & {4, 64} & {6.395}  & {.016} & {.286} \\ 
     Technique  & {2, 32} &  {15.039}  & {\textless .001} & {.485} \\
    Distance  & {2, 32}  & {27.116} & {\textless .001} & {.629} \\
    \bottomrule
\end{tabular}
\vspace{-0.3cm}
% \label{table:1}
\end{table}

\textbf{\revision{Docking} Time:} 
The statistical results are shown in Table~\ref{table_docking_completionTime_result}. A significant interaction effect on the \revision{docking} times between Technique and Target Distance is reported ($p$\textless.001). We find simple effects of Technique with Target Distance at 3m ($p$\textless.001), 6m ($p$=.018), and 9m ($p$=\textless.001). At 3m, PORTAL (M=11.613s) shows a faster \revision{docking} time than LO (M=13.493s, $p$=\textless.001). When Target Distance is 6m and 9m, PORTAL (M\textsubscript{$6m$}=12.812s, M\textsubscript{$9m$}= 10.917s) is faster than HOMER (M\textsubscript{$6m$}=23.995s, $p$=.001 and M\textsubscript{$9m$}=52.492s, $p$\textless.001) and LO (M\textsubscript{$6m$}= 20.617s, $p$=.034 and M\textsubscript{$9m$}=45.916s, $p$\textless.001). Regarding the \revision{docking} time, PORTAL has no simple effect of Target Distance ($p$=.260).

We find a main effect on selection time of Technique ($p$\textless .001). 
PORTAL (M=11.781s) shows the statistically faster \revision{docking} times than HOMER (M=29.993s, $p$\textless.001) and LO (M= 29.363s, $p$\textless.001). 
\vspace{-0.1cm}

\begin{table}[ht]
% \captionof{table}{Table Title} \label{tab:title} 
\caption{\label{table_docking_ErrorDistance_result} \textbf{Task 2: Significant effects on Error Distance}}
\begin{tabular} {m{2.5cm} | m{1cm} | m{1cm}| m{1cm} | m{0.8cm}}
\toprule
    {\textbf{Factor }} & {\textbf{$DOF$}}  & {\textbf{$F$}} & {\textbf{$p$}} & {\textbf{$\eta$\textsubscript{$p$}\textsuperscript{$2$}}}\\ \midrule
       Technique\texttimes Distance & {4, 64} & {2.287}  & {.110} & {.125} \\ 
       Technique  & {2, 32} &  {19.776}  & {\textless .001} & {.553} \\
    Distance  & {2, 32}  & {4.825} & {.018} & {.232} \\
    \bottomrule
\end{tabular}
\vspace{-0.3cm}
% \label{table:1}
\end{table}

\textbf{Error Distance:} 
We report the statistical results in Table~\ref{table_docking_ErrorDistance_result}. 
While no interaction effect is found between Technique and Target Distance, a significant main effect of Technique on error distance is found ($p$\textless.001). 
In pairwise comparisons, PORTAL (M=2.4cm) has  smaller error distances than HOMER (M=3.2cm, $p$\textless.001) and LO (M=3.3cm, $p$\textless.001), while HOMER and LO have similar error distances. At the further analysis for PORTAL, no difference by Target Distance is discovered. 

\subsubsection{Subjective Preferences}

We report statistical differences using the Kruskal-Wallis test at the 5\% significance level based on the survey during Study 1 and post-questionnaire.
The post-hoc test is performed using Dunn's test at the same significance level. We report the results with the median scores and interquartile ranges (IQR) in the form of Median (Q1-Q3). 
% For example, if a questionnaire has a median value of 3 and 2 and 4 for Q1 and Q3, then it is represented in 3 (2-4).
%For example, if a questionnaire has a median, Q1, and Q3 values of 3, 2, and 4 respectively, then it is represented in 3 (2-4).

\textbf{Arm Fatigue:} The significant differences in both Task 1 ($p$\textless.001) and 2 ($p$=.002) are found. 
% HOMER is rated as the technique that caused the least arm fatigue, and its median (IQR) scores are 2 (1-2) and 2 (2-3) for Task 1 and 2 respectively. 
HOMER is rated as the technique that caused the least arm fatigue. Its scores are 2 (1-2) and 2 (2-3) for Task 1 and 2 respectively. 
% It has the significant differences between PORTAL (5 (4-6)), $p$\textless.001) and LO (4 (2-5), $p$=.03) in the selection. 
It has the significant differences between PORTAL (5 (4-6)), $p$\textless.001) and LO (4 (2-5), $p$=.03) in Task 1. 
% For Task 2, PORTAL (3 (2-4)) is ranked the second followed by HOMER, and no significant difference is found between them ($p$=.249). 
For Task 2, PORTAL (3 (2-4)) is ranked the second, but no significant difference between PORTAL and HOMER is found ($p$=.249).

\textbf{Self-Success:}
The highest degree of self-success is reported when the participants used PORTAL in both tasks having 6 (6-6) for Task 1 and 6 (6-7) for Task 2. 
The significant differences between the three in Task 1 ($p$=.046) and Task 2 ($p$\textless.001) are disclosed.
For Task 1, PORTAL has a higher score than HOMER (5 (5-6)) and LO (6 (4-6)), but the pairwise comparison does not report any significant differences. In Task 2, PORTAL shows a statistically higher score than HOMER(4 (4-5), $p$\textless.001) and LO (4 (4-5), $p$\textless.001). The same scores from HOMER and LO are reported. 
% Interestingly, the same scores from HOMER and LO are reported. 

\textbf{Easiness:}
% PORTAL is reported as the easiest technique to select (2 (1-4)) and manipulate (1 (1-2)) the remote targets. 
The participants rate that PORTAL is the easiest for remote object selection (2 (1-4)) and manipulation (1 (1-2)). The significant differences between the three in Task 1 ($p$=.046) and Task 2 ($p$\textless.001) are disclosed. 
The pairwise comparison shows PORTAL is significantly easier for selecting the remote objects than HOMER (4 (3-6), $p$\textless.001) and LO (2 (2-5), $p$=.001). It is also significantly easier than HOMER (5 (4-6), $p$\textless.001) and LO (5 (5-6), $p$\textless.001) in Task 2.

\textbf{Preference: } 15 participants answered that they prefer PORTAL, 5 participants answered HOMER, and only 1 participant answered LO. 
\textbf{P1} who chose PORTAL as his best preference reported that ``PORTAL was the easiest because it was essentially the same thing as controlling an object close to me." %\textbf{P7}, who prefers HOMER the most, reported that PORTAL seems to be better for manipulating remote objects, saying ``I thought that ray-casting was the best for selecting remote objects, but PORTAL was the best for manipulating remote objects because it brought them closer to me." One participant, \textbf{P20}, gave the lowest rating to PORTAL, saying she had trouble opening PORTAL in front of the correct targets. 

\subsection{Discussion on Selection Time} \label{sec_time_for_opening_portals}

\begin{figure}[t]
\centering
\hspace*{-0.5cm}
\includegraphics[width=0.4\textwidth]{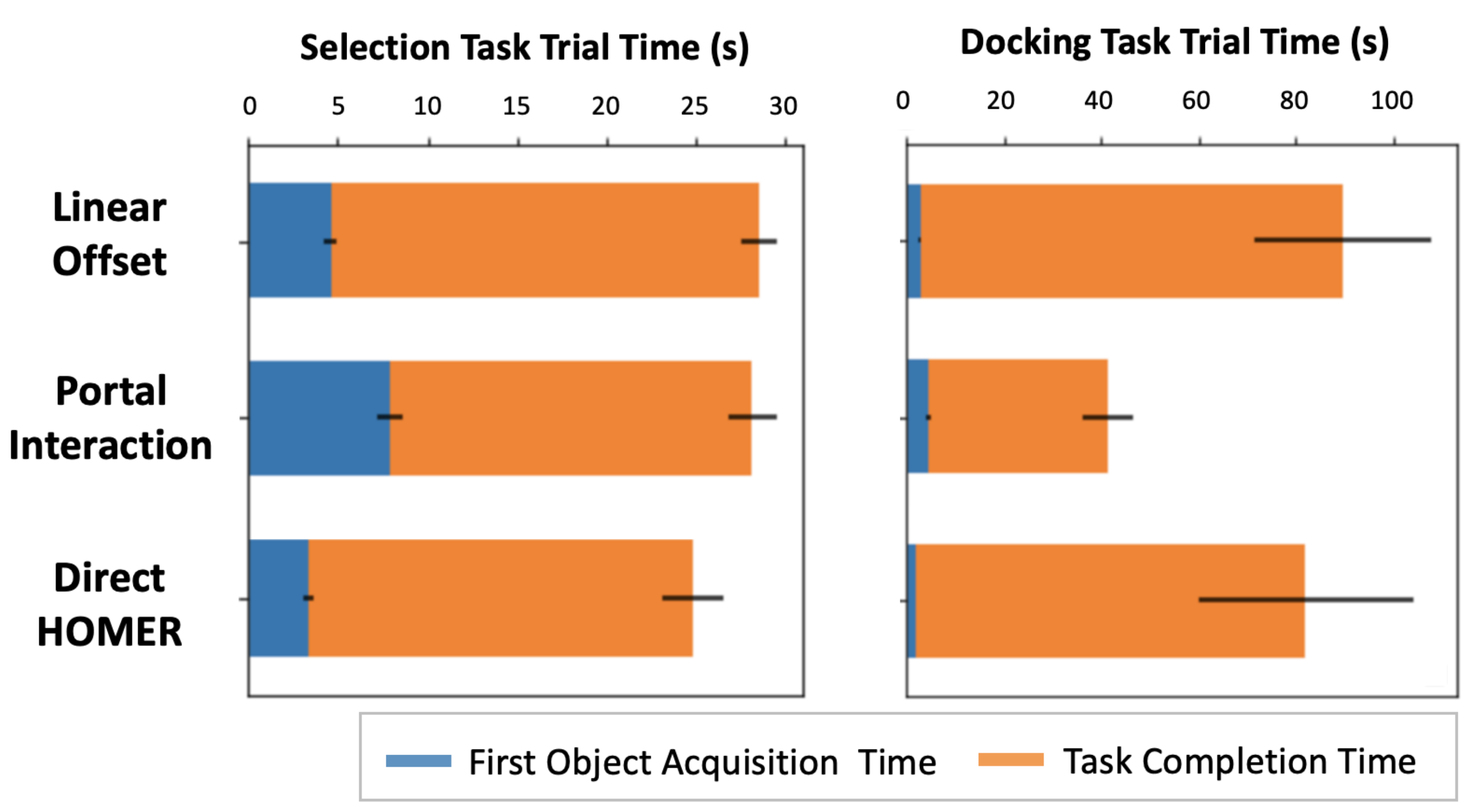}
\caption{Average time consumed for Task 1 and 2. The left and right figures shows the Sphere 0 selection (blue) and task completion times (orange) with 95\% CI error bars.} 
\label{overall_completiontime_trial}   
\vspace{-0.4cm}
\end{figure}

To better understand PORTAL, we discuss the amount of time it took the participant to first hit the target (i.e., Sphere 0 in Fig.~\ref{selection_task_layout} and Docking Object in Fig.~\ref{docking_task_layout}) and the number of times PORTAL was incorrectly opened up to the first hit. The average time for the first target selection and task completion time for each task and technique is shown in Fig.~\ref{overall_completiontime_trial}. 
In Task 1, PORTAL (M=7.753s) takes longer than HOMER (M=3.283s) and LO (M=4.460s) to grab Sphere 0. PORTAL (M=20.302s) has shorter task completion times than HOMER (M=21.436s) and LO (M=23.968s), but its total time increases due to the amount of time spent to open PORTAL correctly. In Task 2, PORTAL (M=4.111s) also takes a longer time than HOMER (M=1.464s) and LO (M=2.367s) to grab Docking Object on average. 
% However, PORTAL still outperforms HOMER and LO in terms of total time to complete the task 2.
However, PORTAL still outperforms HOMER and LO overall. PORTAL took a long time for the first target selection than HOMER and LO. 
% This is because if the user fails to correctly position PORTAL in front of the target object, the user must close it and open a new one. 
This is because if the user fails to correctly position PORTAL, the user must close it and open a new one. In our study, we found that the participants failed 1.593 and 0.889 times to position PORTAL correctly in Task 1 and 2 respectively, taking an average of 2.990s and 2.176s to open PORTAL.

\section{Study 2: PORTAL vs. Teleportation}

% Teleportation is a well-known method to navigate IVEs. 
% The user can interact with out-of-reach objects after teleporting to the nearby position where they located. PORTAL and Teleportation are similar in that both techniques allow the user to interact with the objects right directly in front of the user. However, it is unclear which one could support the user in a more effective manner with better experiences. 
% Therefore, we designed Study 2 to compare PORTAL to Teleportation. Same as Study 1, the functions manipulating PORTAL and sending and receiving an object through the portal are deactivated in Study 2. The study takes approximately 30 minutes and uses the same Apparatus used in  Study 1.

Study 2 (IRB \#12404) is designed to evaluate and compare PORTAL to Teleportation. Teleportation is a well-known navigation method in IVEs. After the teleportation, it is possible that the user can directly interact with the objects without any arm-extension techniques if the target object is located not too low or too high.  %However, it is unclear which one could support the user in a more effective manner with better experiences. It takes approximately 30 minutes.
%  and uses the same Apparatus used in Study 1.

Study 2 uses the same Apparatus and follows the Study 1 procedures. The only difference with Study 1 is an additional questionnaire, Virtual Reality Sickness Questionnaire (VRSQ)~\cite{kim2018virtual} after a participant completes a task. VRSQ evaluates the degree of motion sickness on a 4-point Likert scale (from 0 (None) to 3 (Severe)). Study 2 takes approximately 30 minutes

\subsection{Teleportation}
% We implemented the teleportation technique mimicking the way commonly implemented with the SteamVR plugin\footnote{https://assetstore.unity.com/packages/tools/integration/steamvr-plugin-32647}.
% We implemented the teleportation technique mimicking the way commonly implemented with the SteamVR plugin\cite{STEAMVRPLUGIN}. 
% We implemented the teleportation technique mimicking the way \HDY{supporting from the SteamVR plugin\cite{STEAMVRPLUGIN}. }
% Teleportation, in our implementation, uses the same controller operations as PORTAL, but instead of creating PORTAL it moves to a distant location.
Teleportation, in our implementation, uses the same controller operations as PORTAL. When the user presses the trackpad on the controller, a ray emerges from the virtual hand to the floor in an arc shape. When the user presses the trigger button while pointing a floor position through the ray, it transfers the user to the location with short fade in and out effects.

\subsection{Participants}

% We recruited a total of 15 participants (9 males and 6 females) for Study 2. The average age of the participants is 22.4, ranging from 18 to 28. All the participants were recruited through SONA. The participants were rewarded with the 0.5 SONA point as stipulated in the university SONA policy. 
% \HDY{15 participants were recruited (9 males and 6 females, average age = 22.4, ranging from 18 to 28) through SONA. The participants were rewarded with the 0.5 SONA point.} 
22 participants were recruited (12 males and 10 females, average age = 22.4, ranging from 18 to 28) through SONA. The participants were rewarded with a 0.5 SONA point. All participants were right-handed, and their average arm reach is 54.53cm. All participants have 20/20 (or corrected 20/20) vision and have no impairments in the use of VR devices. Twelve participants reported that they have used VR devices before, and their average self-VR Familiarity score is 3.5 out of 7.0. No participant participated in Study 1.

% \subsection{Procedures}

% Study 2 begins with asking a participant to sign the informed consent form and complete a pre-questionnaire. Following that the instructor briefly introduces Study 2 procedures and two VR techniques (i.e., PORTAL and Teleportation). The participant has a 5-10 minutes training session to be familiar with the techniques. Before wearing the headset, the participant is asked to wear an eye mask (Figure~\ref{vr_devices}) to follow the COVID-19 procedures (IRB \#12404).

% Next, the instructor introduces a task to the participant and instructs him or her to complete it as quickly and accurately as possible. The VR room for the task is the same as Study 1 except the texture on the floor. The floor is represented in the single sky blue color without the white cross checks. Before starting each trial, the participant is asked to move to the center of the VR room where the ``X" mark located. Then the participant can start the trial by clicking the green box.

% When the participant completes the task using one of the techniques, the instructor asks them to take the headset off and to complete two questionnaires. The first questionnaire asks the participants to evaluate the technique's effectiveness on a 7-point Likert scale (from 1 (Very Low) to 7 (Very High)). The second questionnaire is Virtual Reality Sickness Questionnaire (VRSQ)~\cite{kim2018virtual} to evaluate the degree of motion sickness on a 4-point Likert scale (from 0 (None) to 3 (Severe)). When all the tasks are completed, the instructor asks the participant to complete a post-questionnaire. 

\subsection{Task: Selection + Docking}

The task is identical to Task 2 in Study 1. 
% Technique (PORTAL and Teleportation) and Target Distance (3m, 6m, and 9m) are the two independent variables in this study.
Technique (PORTAL and Teleportation) and Target Distance (3m, 6m, and 9m) are the two independent variables. The dependent variables are the selection time, \revision{docking} time, and error distance. The selection time measures how long it takes the participant to first grab the docking object. 
The \revision{docking} time is the time it takes the participant to complete the task after grabbing the object. 
The error distance is a sum of the distances between the vertices of two tetrahedrons. Each participant must complete 54 trials (3 target distances × 2 techniques × 9 trials). The orders of Technique and Target Distance are counterbalanced across all the participants. Our hypotheses are:
% The hypotheses for Study 2 are:
\vspace{-0.1cm}
\begin{enumerate}
% [label=H\arabic*. , wide=0.5em,  leftmargin=*, topsep=0pt, partopsep=0pt]
[label=H\arabic*. ,noitemsep]
\item PORTAL is expected to have faster selection times than Teleportation because Teleportation may require additional movement when the user loses the context of the remote position while PORTAL shows the target in front of the user. 
\item PORTAL and Teleportation is expected to have the same docking completion times because the user can directly interact with the target once they reach targets.
% \item PORTAL and Teleportation have the same amount of errors because both allow the user to directly interact with the target.  
\item PORTAL and Teleportation are expected to have similar errors because both use the same direct interaction technique.  
%allow the user to directly interact with the target.  
\end{enumerate}
\vspace{-0.2cm}

\subsection{Result}
\subsubsection{Quantitative Result}

\begin{figure}[t]
\centering
\includegraphics[width=.5\textwidth]{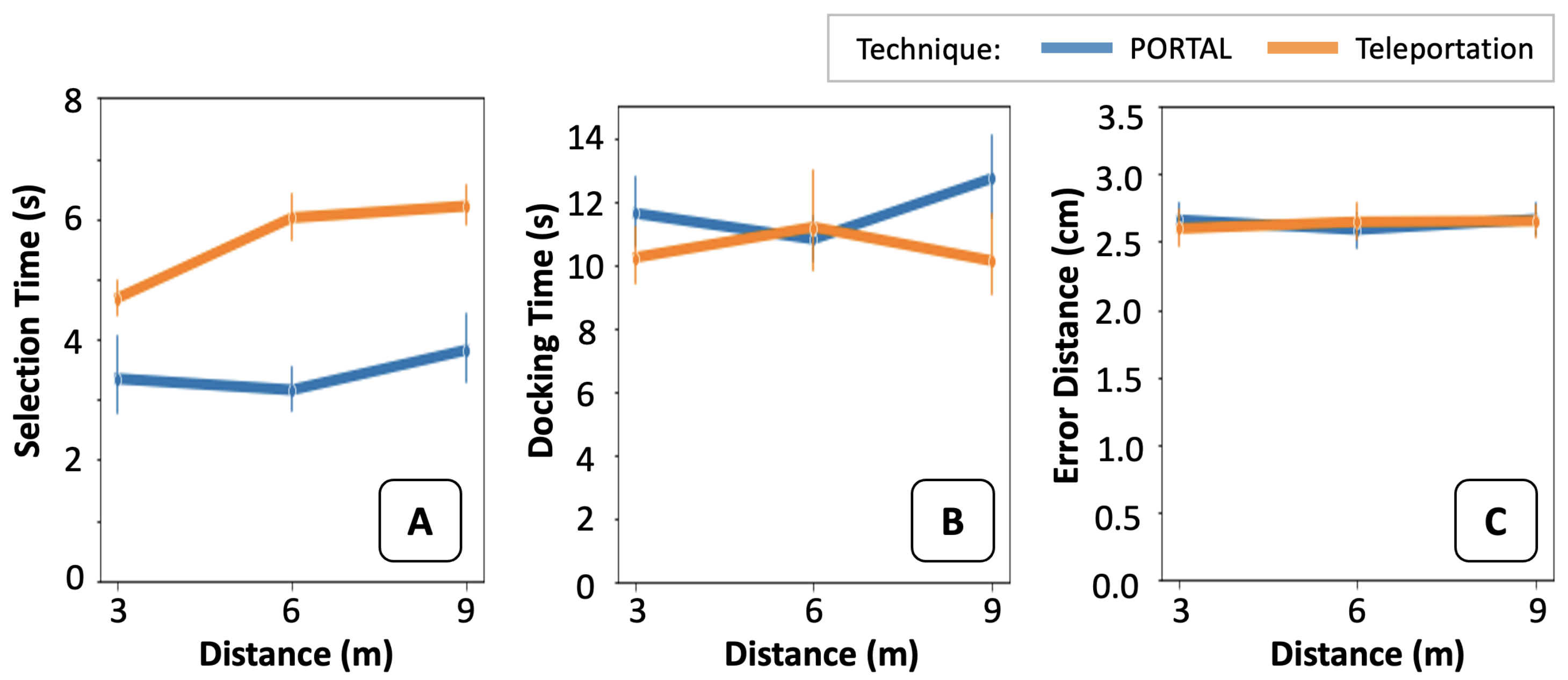}
\caption{
Study 2 Results on the selection times, \revision{docking} times, and error distances with 95\% CI error bars.
} 
\label{study_2_docking_task_results}   
\vspace{-0.3cm}
\end{figure}
The same analytical methods as Study 1 are adopted, the two-way repeated measures ANOVA with Greenhouse-Geisser correction and Tukey's HSD at the 5\% significance level.

\begin{table}[h]
% \captionof{table}{Table Title} \label{tab:title} 
\caption{\label{study2_selection_result_table} \textbf{Study 2:  effects on Selection Time}}
\begin{tabular} {m{2.5cm} | m{1cm} | m{1cm}| m{1cm} | m{0.8cm}}
\toprule
    {\textbf{Factor }} & {\textbf{$DOF$}}  & {\textbf{$F$}} & {\textbf{$p$}} & {\textbf{$\eta$\textsubscript{$p$}\textsuperscript{$2$}}}\\ \midrule
       Technique\texttimes Distance & {2, 42} & {3.441}  & {.052} & {.141} \\ 
       Technique  & {1, 21} &  {28.796}  & {\textless .001} & {.578} \\
    Distance  & {2, 42}  & {8.920} & {\textless .001} & {.298} \\
    \bottomrule
\end{tabular}
% \label{table:1}
\vspace{-0.3cm}
\end{table}

\textbf{Selection Time:} We report the statistical results in Table~\ref{study2_selection_result_table}. 
While no interaction effect is found between Technique and Target Distance, the significant main effects of Technique ($p$\textless.001) and Target Distance ($p$=.001) for the selection time are found. 
PORTAL (M=3.441s) is significantly faster than Teleportation (M=5.633s).

% At the further analysis for PORTAL and Teleportation, each has a significant difference on the selection time by Target distance. PORTAL has the significant effect ($p$\textless.001) between 9m (M=4.432s) and 3m (M=3.463s). Teleportation has the significant differences between 9m and 3m (M\textsubscript{$9m$}=5.792s, M\textsubscript{$3m$}=5.792s, $p$\textless.001) and 6m and 3m (M\textsubscript{$6m$}=4.746s, $p$=.001).
In a more detailed analysis of PORTAL and Teleportation, it was discovered that PORTAL has no significant difference in Target Distance, whereas Teleportation does. Teleportation has the significant differences between 9m and 3m (M\textsubscript{$9m$}=6.207s, M\textsubscript{$3m$}=4.679s, $p$\textless.001) and 6m and 3m (M\textsubscript{$6m$}=6.014s, $p$=.001).

\begin{table}[ht]
% \captionof{table}{Table Title} \label{tab:title} 
\caption{\label{study2_docking_result_table} \textbf{Study 2:  effects on Docking Time}}
\begin{tabular} {m{2.5cm} | m{1cm} | m{1cm}| m{1cm} | m{0.8cm}}
\toprule
    {\textbf{Factor }} & {\textbf{$DOF$}}  & {\textbf{$F$}} & {\textbf{$p$}} & {\textbf{$\eta$\textsubscript{$p$}\textsuperscript{$2$}}}\\ \midrule
       Technique\texttimes Distance & {2, 42} & {2.375}  & {.106} & {.102} \\ 
       Technique  & {1, 21} &  {1.095}  & {.307} & {.050} \\
    Distance  & {2, 42}  & {.364} & {.670} & {.017} \\
    \bottomrule
\end{tabular}
\vspace{-0.3cm}
% \label{table:1}
\end{table}

\textbf{\revision{Docking} Time:} 
The statistical results are represented in Table~\ref{study2_docking_result_table}. It shows no significant effect on Technique, Target Distance as well as no interaction effect between them.

\begin{table}[!h]
% \captionof{table}{Table Title} \label{tab:title} 
\caption{\label{study2_error_result_table} \textbf{Study 2:  effects on Error Distance}}
\begin{tabular} {m{2.5cm} | m{1cm} | m{1cm}| m{1cm} | m{0.8cm}}
\toprule
    {\textbf{Factor }} & {\textbf{$DOF$}}  & {\textbf{$F$}} & {\textbf{$p$}} & {\textbf{$\eta$\textsubscript{$p$}\textsuperscript{$2$}}}\\ \midrule
       Technique\texttimes Distance & {2, 42} & {.474}  & {.608} & {.022} \\ 
       Technique  & {1, 21} &  {.014}  & {.908} & {\textless .001} \\
       Distance  & {2, 42}  & {.292} & {.743} & {.013} \\
    \bottomrule
\end{tabular}
\vspace{-0.4cm}
% \label{table:1}
\end{table}

\textbf{Error Distance:} Table~\ref{study2_error_result_table} shows the statistical results of independent variable on the error distance. We found no significant differences for Technique, Target Distance, and their interaction.

\subsubsection{Subjective Preferences}

% We report the statistical differences in terms of work load and preferences using the Kruskal-Wallis test and Dunn's post-hoc test at the 5\% significance level. We report the median scores and IQR together in the form of Median (Q1-Q3).  
The same Kruskal-Wallis and Dunn's post-hoc tests for Study 1 are used.  
% In addition, we report the one-way ANOVA test result using the VRSQ score at the 5\% significance level.
We also report the one-way ANOVA result for VRSQ.

\textbf{Arm Fatigue:} We found no significant difference between PORTAL (6 (5-6) and Teleportation (6 (6-6)).
%, while Teleportation  receives a slightly higher score than PORTAL (6 (5-6)).

\textbf{Self-Success:} 
% No significant difference between the techniques is found. 
There is no significant difference between PORTAL (6 (5-6)) and Teleportation (6 (6-6)).

%is found. The participants evaluated that they were a little more successful when they use Teleportation (6 (6-6)) than PORTAL (6 (5-6)).

\textbf{Easiness:} Teleportation (2 (1-2.75)) is reported to be easier to use than PORTAL (3 (2-3.75)), but no significant difference is found.

\textbf{Preference: } 5 participants answered that they prefer PORTAL, and 10 participants answered Teleportation.% \textbf{P8} reported he prefers PORTAL saying ``It took me directly to the object with minimal effort," however at the same time he also mentioned that PORTAL was less forgiving when he missed the targets. 
% Most participants who preferred Teleportation reported that Teleportation is intuitive and easier to adjust the misaimed position. 
Most participants who preferred Teleportation reported that it is intuitive and easier to adjust the misaimed position. 
Some participants pointed out that its main difficulty is judging the exact point where teleport to.

\textbf{Motion Sickness: } Motion sickness is a well-known issue which VR users experience when using Teleportation. 
To compare the techniques in terms of motion sickness, we measured the VRSQ score from the participants' responses following \ref{VRSQ_eq}. 
\textbf{\textit{i}} refers the question number in the VRSQ questionnaire, and \textbf{\textit{s(i)}} indicates each score ranging from 0 to 3.    
\vspace{-0.2cm}
\begin{equation} \label{VRSQ_eq}
\vspace{-0.2cm}
VRSQ Score=((\sum_{i=1}^{4} s(i)) /12 *100 + (\sum_{i=5}^{9} s(i)) /15 *100 )/2
\end{equation}
% \vspace{-0.1cm}
The participants reported that they felt less motion sickness for PORTAL (M=5.11) than Teleportation (M=6.85). However, no significant difference is found.

\section{Discussion}
\label{section_discussion}
% We evaluate PORTAL by the two user studies compared to the existing techniques designed for interacting with remote objects in IVEs. 
% In Study 1, we conducted two tasks to evaluate three interaction techniques (PORTAL, HOMER, and LO).
In Study 1, we conducted two tasks to evaluate PORTAL, HOMER, and LO. 
% Figure~\ref{selection_task_results} clearly shows that the performances of HOMER and LO are getting worse as the target distance is increased, while PORTAL is not affected by the distance in the previous section (H1, H3). 
Fig.~\ref{selection_task_results} clearly shows the Task 1 results. The performances of HOMER and LO are getting worse as the target distance is increased, while PORTAL is not affected by the distance (H1, H3). In addition, PORTAL outperforms HOMER and LO in terms of accuracy at all distances (H2). HOMER adopts ray-casting for target selection, and it would be the reason for its lower performance as ray-casting is easily influenced by the controllers' jitter when targeting further objects~\cite{accot2001scale, benko2006precise}, resulting in incorrect target selection (known as Heisenberg effect ~\cite{bowman2001using}).
% Though the target size and distances in Study 1 differ from Poupyrev and Ichikawa's work~\cite{poupyrev1999manipulating}, Task 1 for HOMER shows the same result with their finding, in which HOMER suddenly dropped its performance when high accuracy is required. 
% Though the target size and distances differ from Poupyrev and Ichikawa's work~\cite{poupyrev1999manipulating}, Task 1 for HOMER shows the same result with their finding, in which HOMER suddenly dropped its performance when high accuracy is required. 
Though the target size and distances differ from Poupyrev and Ichikawa's work~\cite{poupyrev1999manipulating}, HOMER shows the same result as their finding, in which HOMER suddenly dropped its performance when high accuracy is required. 
% Because PORTAL also uses ray-casting as its selection tool, it suffers from the same problem as P20 in Study 1 pointed out.
As PORTAL also uses ray-casting, it suffers from the same problem for the first target hit.
% We will discuss more about this in the following sub-section.

In Task 2, PORTAL at any distance has better performances than HOMER and LO. 
% The results show that PORTAL has the potential to help the user manipulate the remote objects by providing proper depth perception, supporting H1 and H2. 
The results show that PORTAL successfully helps the user manipulate the remote objects by providing proper depth perception, supporting H1 and H2. 
% We also find that PORTAL has constant performance by the distance (H3). 
PORTAL has constant performance by the distance (H3). We find no different performance between HOMER and LO. It is not a surprising result as the same linear offset is applied to both once the user grabs remote objects. 

In Study 2, we compared PORTAL and Teleportation. The results show PORTAL is significantly faster than Teleportation to access the target objects. It supports our H1. 
We also discovered that the target distances do not affect the selection time when using PORTAL. It is the consistent result with the Task 1 results in Study 1. 
% PORTAL took a longer time when the distance is increased. We can conclude that the participants had trouble with opening PORTAL when the target object appeared smaller along the target distances. 
% PORTAL and Teleportation show no significant differences for the docking time and errors, supporting H2 and H3. 
PORTAL and Teleportation show no significant differences in the docking time and errors (H2, H3). 
Though PORTAL shows slightly better performance and no significant differences in the surveys, we found more participants preferred Teleportation. One reason could be familiarity. Teleportation is used in Vive and Oculus UIs, so the participants are more familiar with it than PORTAL.
% Another reason would be that, with Teleportation, the user can physically move around the target object to see the other side.
Another reason would be that, with Teleportation, the user can physically move around the target object to adjust the misaimed position. It also allows the user to easily navigate remote places around the target. PORTAL supports this as well, but in limited angles without changing its position and orientation.

\subsection{PORTAL Potential Extensions}

In this subsection, we discuss the applicable issues using PORTAL.

% \textbf{Occlusion: } 
% Occlusion is a challenging issue for target selections in IVEs, occurred when an object is blocking another object from the user's view. We expect that PORTAL has a potential to ease the occlusion problem by controlling PORTAL's position and orientation. There are two possible approaches. The first approach is to specify the direction of the secondary portal in relation to a target when opening PORTAL. The primary portal then projects not only the front view but also the non-frontal view such as the top, left, or right side of the target. This allows the user to interact with the target from various angles, thus will solve the occlusion problem. By including a step allowing the user to specify the direction of the secondary portal, the user can explore occluded objects behind the target as well as interact with them. 

% The second approach is to manipulate the position and orientation of the portal objects after they are created. As we described in section~\ref{subsection_portal_interaction}, the user can grab and move PORTAL, and it affects the projected view on the primary portal. By manipulating the primary portal, the user can navigate remote places around the target, and it results to ease the occlusion problem as well. It is also possible that the user can use PORTAL to see through a wall by creating a portal right in front of the wall and move the portal to find objects behind the wall. For future work, we are planning to conduct user studies to evaluate how effectively PORTAL can help the user with the occlusion in IVEs. 

\textbf{Occlusion: } It is a challenging issue for target selections in IVEs, occurred when an object is blocking another object from the user's view. PORTAL has the potential to ease the problem by controlling PORTAL's position and orientation. As we described in section~\ref{subsection_portal_interaction}, the user can relocate PORTAL, and it affects the projected view on the primary portal. It allows the user to interact with the target from various angles, thus this will solve the occlusion problem. 

% \textbf{Multiple Portals Interaction:}
\textbf{Multiple Portals:}
% Current work allows the user to create a single set of PORTAL. 
% However, by placing more than one set, it may provide additional interactions for the user. 
Placing more than one set of PORTAL may provide multi-space interactions. The user can move an object among multiple remote scenes without navigation like \cite{kiyokawa2005tunnel}. It could be achieved by getting the object to the user's side first through the first portal and putting it into the next portal.
% For example, if the IVE has three rooms (room A, B, and C), the user can move an object among three different rooms. In this case, the user in room A can open two PORTALs for room B and C for the multi-space interaction.

PORTAL can be used for multi-scale VEs \cite{cho2014evaluating, cho2017multi} for interaction and navigation through multiple VR scenes with different scales. For example, in an application like Google Earth, the user can bring some objects from home to another place, city, country, or planet.

\textbf{More Interaction Techniques with PORTAL: } 
% \textbf{Other Interactions: } 
% We evaluated PORTAL for one hand 6-DOF interactions with remote objects in this paper. 
PORTAL has the advantage of allowing the user to interact with objects with both hands. 
% This enables us to easily use other uni- and bi-manual techniques to support simultaneous 6- and 7-DOF (6-DOF with scale) manipulation, such as Grab-and-Scale \cite{cutler1997two} and Spindle+Wheel~\cite{cho2015evaluation} as well as multi-modal interfaces (e.g., voice and gaze) and natural user interfaces (e.g., Air TRS \cite{de2013mockup}). 
This enables us to easily use other uni- and bi-manual techniques to support simultaneous 7-DOF (6-DOF with scale) manipulation, such as Grab-and-Scale \cite{cutler1997two} and Spindle+Wheel~\cite{cho2015evaluation}.
% as well as natural user interfaces (e.g., Air TRS \cite{de2013mockup}). 

PORTAL can be used as a navigation technique by moving the user's entire body into the portal~\cite{liu2018increasing}. It also has a potential to support the user's gradual transition between VR scenes~\cite{steinicke2009transitional}. 

% \needToClarification {\textbf{AR/MR:} PORTAL is considered and designed for VR in this work. However, it has a potential to be applied in an AR and MR space. }

% \textbf{AR/MR:} PORTAL has the potential to be used in AR and MR applications. 
% % However, locating a secondary camera and projecting its view on PORTAL in an AR or MR environment would be challenging because it would have to take into account both 3D virtual and real-world objects.
% \HDY{However, locating a second camera and projecting its view on PORTAL in AR and MR would be challenging because it would take both 3D virtual and real-world objects into account. }

\subsection{Limitations and Future Work}

\revision{Our studies show that PORTAL is a suitable tool to control the exact location and orientation of distant objects. However, it may not be suitable in the following scenarios: 1. A target object looks too small from the user’s location; 2. The object looks too big, and thus its local contexts are not clearly visible within a portal; and 3. The tasks require to adjust the target object's position in a larger space. For these limitations, we suggest the following future work.}
% PORTAL was successfully assisting the participants in interacting with remote objects, but we found some limitations. 
% We discuss the limitations and potential methods to improve PORTAL. 
% We discuss the limitations and \HDY{potential solutions for PORTAL.}
% In this subsection, \HDY{we discuss the potential methods to improve it. }

\textbf{Selection Tool:} 
Ray-casting caused the incorrect PORTAL creation. Volumetric selection tools such as cone shape metaphor~\cite{liang1994jdcad, steed2006towards, olwal2003senseshapes} and heuristic approaches~\cite{steinicke2006object, de2005intenselect} could be alternatives. 
Adopting these methods for PORTAL \revision{would solve} this issue.
% Adopting these methods for PORTAL is expected to reduce this issue.

\textbf{PORTAL Size: } A visual discontinuity between the remote and local positions occurs at the edge of PORTAL. It could affect the user to understand contexts of remote and local positions. Its impact, however, is unclear. 
\revision{Our future research will focus on comparing different PORTAL sizes following target object sizes to determine an optimal PORTAL size.}  
% Our future research will focus on determining its impact and optimal PORTAL sizes, as well as evaluating the function of bringing the object to the user's side and sending an object to the remote side via PORTAL. 
%In addition, modeling PORTAL performance as a function of target sizes are also our future research.

\textbf{PORTAL Placement in Arbitrary Positions: }
The current work requires a target to open PORTAL.
% Where to open a portal without targets is challenging. 
% We can use a similar approach as teleportation by creating multiple waypoints in an IVE. 
% It may require a different interaction technique than ray-casting to create PORTAL. 
Placing it in random places where the user intends in the IVE is our future research. 

\section{Conclusion}
\label{section_conclusion}
% In this paper, we introduced PORTAL, a novel method for interacting with remote objects in IVEs through direct hand access and control. 
In this paper, we introduced PORTAL for interacting with remote objects in IVEs through direct hand access and control. PORTAL is an interactive widget leveraging the principle of secondary view interaction to explore remote views without navigation. It allows the user to interact with remote objects by providing an illusion that the remote objects are located within-reach distance. We evaluated PORTAL and reported its performance from the two empirical studies. % that include selecting and manipulating objects located at the three distances (3m, 6m, and 9m) in comparison to the existing techniques, direct HOMER, Linear Offset, and Teleportation. 
In Study 1, Task 1 shows that PORTAL has the fastest selection time in 9m while direct HOMER is faster in 3m and 6m. In addition, PORTAL has the lowest error rate regardless of the depth. 
% Task 2 shows that PORTAL has the fastest and most accurate performance in comparison to direct HOMER and Linear Offset.
Task 2 shows that PORTAL has the fastest and most accurate performance. In Study 2, PORTAL shows the faster selection time and has the same performance on manipulation time and accuracy compared to Teleportation. Based on our findings, we discussed the potential directions and improvements for future research that PORTAL could be extended beyond its limitations.

\bibliographystyle{ACM-Reference-Format}
\bibliography{Contents/0_main.bib}

\end{document}